# Managing engineering systems with large state and action spaces through deep reinforcement learning

C.P. Andriotis, K.G. Papakonstantinou

Department of Civil & Environmental Engineering
The Pennsylvania State University, University Park, PA, 16802, USA

## Abstract

Decision-making for engineering systems can be efficiently formulated as a Markov Decision Process (MDP) or a Partially Observable MDP (POMDP). Typical MDP and POMDP solution procedures utilize offline knowledge about the environment and provide detailed policies for relatively small systems with tractable state and action spaces. However, in large multi-component systems the sizes of these spaces easily explode, as system states and actions scale exponentially with the number of components, whereas environment dynamics are difficult to be described in explicit forms for the entire system and may only be accessible through numerical simulators. In this work, to address these issues, an integrated Deep Reinforcement Learning (DRL) framework is introduced. The *Deep Centralized Multi-agent Actor Critic* (*DCMAC*) is developed, an off-policy actor-critic DRL approach, providing efficient life-cycle policies for large multi-component systems operating in high-dimensional spaces. Apart from deep function approximations that parametrize large state spaces, DCMAC also adopts a factorized representation of the system actions, being able to designate individualized component- and subsystem-level decisions, while maintaining a centralized value function for the entire system. DCMAC compares well against Deep Q-Network (DQN) solutions and exact policies, where applicable, and outperforms optimized baselines that are based on time-based, condition-based and periodic policies.

*Keywords*: multi-component deteriorating systems; inspection and maintenance; partially observable MDPs; multi-agent stochastic control; large discrete action spaces; deep reinforcement learning

## 1. Introduction

Efficient management of structures and infrastructure is an ever timely issue of paramount importance, aiming at proper inspection and maintenance policies, able to handle various stochastic deteriorating effects and suggest optimal actions that serve multi-purpose life-cycle objectives. Optimality at a single-component level, although often well defined, is not sufficient to ensure optimality at the system level, where interactions of different components may be able to prompt or mitigate failure events and often unveil uncommon performance dynamics. Similarly, as a result of structural dependence, combinations of certain component states may suggest diverse maintenance and inspection actions at different parts of the system, potentially encouraging decisions that would be otherwise largely sub-optimal under single-component optimality conditions. Comprehensive control of large systems, however, is not always amenable to conventional computational concepts and can become particularly challenging as a result of the immense number of possible state and action combinations involved in the decision process. This paper addresses the need for intelligent and versatile decision-making frameworks that have the capacity to operate in complex high-dimensional state and action domains, and support analysis, prediction, and policy-making in systems with multiple dissimilar and interdependent constituents.

Multiple solution procedures exist in the literature for planning maintenance and inspection policies for structures and systems, which can be discerned, among others, in relation to their optimization procedure, the defined objective function and decision variables, the deteriorating environment dynamics, the available action-set at each decision step, and the accuracy of information gathered from data [1]. Regardless of their individual specifications, a central feature in existing methods searching for optimum maintenance and inspection actions over the planning horizon is the utilization of time- and threshold-based approaches. Thresholds are typically chosen in reference to certain metrics of interest, such as system condition, reliability, or risk, whereas times are usually determined through periodic or non-periodic assumptions. Fixed reliability thresholds have been utilized to optimize visit times for maintenance and inspection, e.g. in [2, 3]. Bayesian theory and fixed reliability-based thresholds have been combined, to determine the inspection times over the life-cycle, for given repair strategies [4]. Optimization solution schemes for determining proper inspection times have been also formulated within the premises of Bayesian networks, similarly utilizing risk-based thresholds and given condition-based criteria for repairs [5, 6]. Other formulations emanate from renewal theory, usually proposing predefined condition-based thresholds for replacements [7], to determine periodic policies that optimize time intervals between inspections. In the most general approach, both optimal repair thresholds and inspection times or time intervals are sought, e.g. in [6, 8]. Overall, decisions are derived through constrained or unconstrained static optimization formulations, typically supported by either gradient-based optimization schemes, or evolutionary and simulation-based algorithms that are shown, in some cases, to be able to handle more efficiently the combinations of discrete and continuous decision variables in multi-objective problems, e.g. [9, 10]. Nonetheless, as the decision problem becomes more complex, pertaining to long-term sequential decisions and multi-component domains, efficient solutions can become particularly challenging, often requiring adoption of convenient modeling simplifications, usually at the expense of accuracy and solution quality.

Markov Decision Processes (MDPs) provide a solid mathematical framework for optimal sequential decision-making and have a long history of research and implementation in infrastructure systems applications [11, 12], as they provide strong global optimality guarantees for long-term objectives, either through dynamic programming solution schemes or linear programming formulations [13]. A number of infrastructure asset management decision-support systems have, therefore, embraced MDP principles, suggesting implementation of discrete Markovian dynamics for the characterization of system states in time, e.g. [14], whereas the appropriateness and efficiency of Markovian models for predicting the evolution of damage states in discrete spaces has been demonstrated in a variety of structural settings and for different exogenous stressors [15, 16, 17, 18, 19]. Despite their unique qualities, MDPs are not always suitable for decision-making in engineering systems, as they are formulated under the assumption of complete information and error-free observations, thus requiring knowledge of the exact state of the system prior to an action at every decision step.

A natural extension to MDPs that relaxes the limitations of complete information are Partially Observable Markov Decision Processes (POMDPs), which have also been implemented in a number of engineering and structural engineering applications [20, 21, 22, 23]. As a result of the partially observable environments in POMDPs, the obtained policies are conditioned on beliefs over the system states, which essentially encode the entire history of states, actions and observations through Bayesian updates. Beliefs, however, form continuous state-spaces in this case that complicate the solution procedure. In response to this, point-based algorithms have been introduced, primarily in the fields of machine learning, artificial intelligence, and robotics [24, 25, 26] and recently in structural applications [27, 28, 29], showcasing substantial capabilities by virtue of their apposite belief space exploration heuristics and synchronous or asynchronous recursive Bellman backups. Combinations of MDP and POMDP environments can be also taken into account, through Mixed Observability Markov Decision Processes (MOMDPs), which are supported by specialized algorithms that can reformulate the problem through suitable factorized state representations [30].

Conventional MDP and POMDP formulations suit well certain low-dimensional domains, typically corresponding to system components or simple systems with a limited number of system states and actions. However, they can often become impractical when large-scale multi-component domains are considered, as system state and action spaces scale exponentially with the number of components in the most detailed maintenance and inspection scenarios, Markovian transition matrices become extremely large, and computational complexity of action-value function evaluations per decision step severely deteriorates. For example, a stationary system with 20 components, 5 states and 5 actions per component is fully described by nearly $10^{14}$ states and actions! This issue renders the problem practically intractable by any conventional solution scheme or advanced MDP or POMDP algorithm, unless domain knowledge and simplified modeling can possibly suggest drastic state and action space moderations. A straightforward modeling approach facilitating convenient state and action spaces reductions is to exploit similarity of components. This assumption may suffice in systems where components are highly homogeneous and structurally independent, e.g. wind farms where turbines can be assumed to share similar properties with negligible component interactions [31]. In other cases, it is feasible to properly engineer macro-states and -actions in order to achieve practical problem-specific state and action space reductions, e.g. in [32], or to take advantage of the underlying structural properties of the state space in order to construct compressed representations that enable the applicability of traditional MDP and POMDP solvers [33].

Reinforcement Learning (RL) is theoretically able to alleviate the curse of dimensionality related to the state space, either under model-free approaches that do not utilize prior offline environment information on transition dynamics, or model-based approaches that also try to learn the underlying transition model of the environment, or hybrid approaches thereof [34, 35]. In RL, the decision-maker, also called agent, does not synchronously update the value function over the entire state space, but merely conducts updates at states that are visited while probing the environment, without the need for a priori explicit knowledge of the entire environment characteristics. Classical RL techniques have also been implemented for maintenance of engineering systems, providing approximate solutions in various settings, e.g. in [36, 37, 38]. Unfortunately, RL exhibits several limitations in practice when deployed in high-dimensional and complex stochastic domains, mainly manifesting algorithmic instabilities with solutions that significantly diverge from optimal regions, or exhibiting slow value updates at infrequently visited states. However, with the aid of deep learning, RL has been recently driven to remarkable breakthroughs, which have signaled a new era for autonomous control and decision-making. Deep Reinforcement Learning (DRL) has brought unprecedented algorithmic capabilities in providing adept solutions to a great span of complex learning and planning tasks, even outperforming human control at domains that were traditionally dominated by human experts [39, 40]. Similarly to the great progress that deep learning has enabled in machine learning and artificial intelligence [41], DRL agents are capable of discovering meaningful parametrizations of immense state spaces through appropriate deep neural network architectures, and learning near-optimal control policies by interacting with the environment.

In this work, we investigate and develop DRL architectures and algorithms, favorably tailored to stochastic control and management of large engineering systems, and to the best of our knowledge this is the first work casting similar decision-making problems in a DRL framework. We discuss and examine the performance and appropriateness of both Deep Q-Network (DQN) [39] and deep policy gradient architectures, and introduce the *Deep Centralized Multi-agent Actor Critic* (*DCMAC*) approach, along the lines of deep off-policy actor-critic algorithms with experience replay [42, 43]. DCMAC provides efficient life-cycle policies in otherwise practically intractable problems of multi-component systems operating in high-dimensional state and action spaces, and is favorably constructed for providing comprehensive individualized component- and subsystem-level decisions, while maintaining and improving a centralized value function for the entire system. In DCMAC, two deep networks (actor and critic) co-exist and are trained in parallel, based on environment signals and replayed transitions that are retrieved from the agent's experience. DCMAC can handle complex systems under complete and incomplete information, in both MDP and POMDP settings, and is thoroughly evaluated in a series of diverse numerical examples against DQN solutions and exact policies, where applicable, as well as various optimized baseline policies.

## 2. Markov Decision Processes

MDPs provide solutions for optimal sequential decision-making in stochastic environments with uncertain action outcomes and exact observations. The environment, *E*, unfolds in discrete steps and is defined by a finite set of states, *S*, a stochastic interstate transition model, a reward function, *r*, and a finite set of actions, *A*. At each decision step *t*, the decision-maker (called agent in the remainder) observes the current state $s_t \in S$, takes an action $a_t \in A$, receives a reward as a result of this state and action, $r(s_t, a_t)$, and proceeds to the next state $s_{t+1} \in S$, according to the underlying Markovian transition probabilities, $p(s_{t+1} | s_t, a_t)$. It is, therefore, assumed that the current state and selected action are sufficient statistics for the next state, regardless of the entire prior history of state and action sequences. It is also important to note here that the Markovian property is not restrictive in any sense, since environments that do not directly possess it can be easily transformed to Markovian ones through state augmentation techniques, e.g. [21].

The state-dependent sequence of actions defines the agent's policy, $\pi$. Agent's policy can be either deterministic, $\pi(s_t): S \to A$, mapping states to actions, or stochastic, $\pi(a_t | s_t): S \to P(A)$, mapping states to action probabilities. For a deterministic policy, $\pi$ is a single value, given $s_t$. In the case of discrete actions, a stochastic policy $\pi$, given $s_t$, is a vector defining a probability mass function over all possible actions, whereas for continuous actions, the policy is a probability density function. To keep notation succinct and general, all policies are shown in non-vector notation in the remainder of this work. Policy, $\pi$, is associated with a corresponding total return, $R_t^\pi$, which is the total reward collected under this policy, from any time step *t* to the end of the planning horizon *T*:

$$R_t^\pi = r(s_t, a_t) + \ldots + \gamma^{T-t} r(s_T, a_T) = \sum_{i=t}^{T} \gamma^{i-t} r(s_i, a_i) \tag{1}$$

where $\gamma$ is the discount factor, a positive scalar less than 1, indicating the increased importance of current against future decisions. Considering the two extreme cases, $\gamma = 0$ indicates that only the decision at time *t* matters, whereas with $\gamma = 1$ every decision up to time *T* is equally important. The total return in Eq. (1) is a random variable, as state transitions and, potentially, policies are stochastic. Conditioning the total return on the current state-action pair, $s_t, a_t$, the action-value function, $Q^\pi$, is defined as the expected return over all possible future states and actions:

$$Q^\pi(s_t, a_t) = \mathbb{E}_{s_{i>t} \sim E, a_{i>t} \sim \pi}\left[R_t^\pi | s_t, a_t\right] \tag{2}$$

Using Eqs. (1) and (2), the action-value function can be defined through the following convenient recursive form for any given policy, $\pi$:

$$Q^\pi(s_t, a_t) = r(s_t, a_t) + \gamma \mathbb{E}_{s_{t+1} \sim E, a_{t+1} \sim \pi}\left[Q^\pi(s_{t+1}, a_{t+1})\right] \tag{3}$$

The value function, or total expected return from state $s_t$, for policy $\pi$, is defined as the expectation of the action-value function over all possible actions at the current step:

$$V^\pi(s_t) = \mathbb{E}_{a_t \sim \pi}\left[Q^\pi(s_t, a_t)\right] \tag{4}$$

Under standard conditions for discounted MDPs, out of all possible policies there exists at least one deterministic policy that is optimal, maximizing $V^\pi(s_t)$ [44]. For a deterministic policy, with a given model of transitions $p(s_{t+1} | s_t, a_t)$ and the aid of Eqs. (3) and (4), the optimal action-value and value functions, $Q(s_t, a_t)$ and $V(s_t)$, respectively, follow the Bellman equation [45]:

$$\begin{aligned} V(s_t) &= \max_{a_t \in A}\{Q(s_t, a_t)\} \\ &= \max_{a_t \in A}\left\{r(s_t, a_t) + \gamma \mathbb{E}_{s_{t+1} \sim E}\left[\max_{a_{t+1} \in A} Q(s_{t+1}, a_{t+1})\right]\right\} \\ &= \max_{a_t \in A}\left\{r(s_t, a_t) + \gamma \mathbb{E}_{s_{t+1} \sim E}\left[V(s_{t+1})\right]\right\} \\ &= \max_{a_t \in A}\left\{r(s_t, a_t) + \gamma \sum_{s_{t+1} \in S} p(s_{t+1} | s_t, a_t) V(s_{t+1})\right\} \end{aligned} \tag{5}$$

Eq. (5) describes the standard MDP objective which is typically solved using value iteration, policy iteration, or linear programming formulations [13]. A concise MDP presentation can be also seen in [21].

### 2.1. Partially Observable MDPs

POMDPs extend MDPs to partially observable environments. Unlike MDPs, in POMDPs the agent cannot observe the exact state of the system, $s_t$, at each decision step, but can only form a belief about the states, $\mathbf{b}_t$, which is a probabilistic distribution over all possible states. More specifically, starting at belief $\mathbf{b}_t$, the agent takes action $a_t$ and the system transitions to its new state $s_{t+1}$, which is, however, hidden for the agent. The agent instead receives an observation $o_{t+1} \in \Omega$, and subsequently forms the next belief, $\mathbf{b}_{t+1}$, according to a known model of conditional probabilities $p(o_{t+1} | s_{t+1}, a_t)$ and the Bayesian update [21]:

$$\begin{aligned} b(s_{t+1}) &= p(s_{t+1} | o_{t+1}, a_t, b_t) \\ &= \frac{p(o_{t+1} | s_{t+1}, a_t)}{p(o_{t+1} | \mathbf{b}_t, a_t)} \sum_{s_t \in S} p(s_{t+1} | s_t, a_t) b(s_t) \end{aligned} \tag{6}$$

where probabilities $b(s_t)$, for all $s_t \in S$, form the belief vector $\mathbf{b}_t$ of length $|S|$, and the denominator of Eq. (6), $p(o_{t+1} | \mathbf{b}_t, a_t)$, is the standard normalizing constant.

As implied by the updating scheme in Eq. (6), in a POMDP setting the agent takes actions and receives observations that change its perception for the system state probabilities, moving from one belief to another. As such, beliefs can be seen as alternative states of this environment, and POMDPs can be accordingly regarded as belief-state MDPs. Following this statement, the total expected return takes an expression similar to Eq. (5):

$$V(\mathbf{b}_t) = \max_{a_t \in A}\left\{ \sum_{s_t \in S} b(s_t) r(s_t, a_t) + \gamma \sum_{o_{t+1} \in \Omega} p(o_{t+1} \mid \mathbf{b}_t, a_t) V(\mathbf{b}_{t+1}) \right\} \quad (7)$$

Despite this convenient conceptual consistency with MDPs, POMDPs are not as easy to solve. Note that the new belief-state space is not discrete but continuous, forming a $(|S|-1)-$ dimensional simplex. It turns out, however, that in this case, value function, $V$, is convex and piecewise linear, and can be precisely represented with a finite set $\Gamma$ of affine hyperplanes, known as $\alpha$ - vectors, over the continuum of points of the belief space [46]:

$$V(\mathbf{b}_t) = \max_{\boldsymbol{\alpha} \in \Gamma}\left\{ \sum_{s_t \in S} b(s_t) \alpha(s_t) \right\} \quad (8)$$

A suitable set $\Gamma$ and its corresponding $\alpha$-vectors, properly supporting the belief space, is what point-based algorithms seek to determine in order to solve the problem. Further details and insights about point-based methods can be found in [47], whereas detailed comparisons and implementations in relation to maintenance and inspection planning problems for deteriorating structures are provided in [29].

### 2.2. Reinforcement Learning

As shown in Eqs. (5)-(7), typical approaches to solving MDPs and POMDPs presume offline knowledge of explicit probabilistic models for the environment of the entire system. This requirement can be eliminated by RL approaches which make no use of offline domain knowledge and learn to act optimally by directly interacting with *E*. Typical solution techniques implement temporal difference updates on the action-value function, $Q$, based on samples from *E*, such as:

$$\begin{aligned} &Q(s_t, a_t) \leftarrow Q(s_t, a_t) + \eta\left(Q(s_t, a_t) - y_t\right) \\ &y_t = r(s_t, a_t) + \gamma \max_{a_{t+1} \in A} Q(s_{t+1}, a_{t+1}) \end{aligned} \quad (9)$$

where $\eta$ is an appropriate scalar learning rate. Temporal difference learning is central in many RL algorithms and there exists a number of alternative formulations employing Eq. (9), either in its fundamental form or variations of it that depend on the specifications of the training scheme [48]. Nonetheless, all approaches can be largely categorized in two major groups, namely on-policy and off-policy algorithms. In on-policy RL, the agent interacts with the environment while executing the actual policy it learns, e.g. SARSA algorithm [49], whereas in off-policy RL the agent learns the optimal policy while executing a different behavior policy, e.g. Q-learning algorithm [50]. The learning problem in RL can be also approached by directly updating the policy, $\pi$, based on the policy gradient theorem [34]. These methods adopt a parametric approximation of the policy and, as such, are covered by the description of DRL and deep policy gradient methods, which are discussed in the next section.

## 3. Deep Reinforcement Learning

As a result of discrete state and action spaces, action-value functions, can take cumbersome tabular forms, with their updates proceeding asynchronously at visited states or group of states during the training steps, as in Eq. (9). Similarly, policy functions and value functions have vectorized representations according to the cardinality of *S*. It is, therefore, clear that adequate evaluation of such functions is particularly hard in complex environments with extremely large state spaces, whereas continuous spaces have similar issues and are not always easy to describe through consistent mathematical formulations. The key idea of DRL is to utilize deep neural networks as function approximators to efficiently parametrize the state space, thus essentially providing arbitrarily accurate proxies of the original functions, such as:

$$F \simeq F(\cdot \mid \boldsymbol{\theta}^F) \quad (10)$$

where $F$ is one of the previously defined functions $Q^\pi, Q, V^\pi, V, \pi$, and $\boldsymbol{\theta}^F \in \Theta$ are real-valued vectors of parameters. Thereby, the whole problem of determining values at each point of a high-dimensional space, reduces to determining a number of parameters, with $|\Theta| \ll |S \times A|$. We succinctly present below the two core DRL approaches, which are also employed in this work and form the basis for our proposed DCMAC approach and the pertinent algorithmic and numerical implementations that follow.

### 3.1 Deep Q-Networks

The most straightforward DRL approach is to directly parametrize the Q-function with a deep network, and integrate it in an off-policy Q-learning scheme. This is a powerful methodology originally presented in [51], where a Deep Q-Network (DQN) is trained in a suite of Atari games. The general DQN concept is illustrated in Fig. 1(a). State $s_t$ is introduced as input to a deep neural network, with an appropriate number of hidden layers and nonlinear unit activations, which output an approximation of the action-value function $Q\left(s_t, a_t \mid \boldsymbol{\theta}^Q\right) \in \mathbb{R}^{|A|}$. The objective function that is minimized during training to determine $\boldsymbol{\theta}^Q$ is given by the loss function:

$$L_Q\left(\boldsymbol{\theta}^Q\right) = \mathbb{E}_{s_t \sim \rho, a_t \sim \mu}\left[\left(Q\left(s_t, a_t \mid \boldsymbol{\theta}^Q\right) - y_t\right)^2\right] \tag{11}$$

where $\mu$ is the behavior policy of off-policy learning, with $\mu \neq \pi$, $\rho$ is the limiting distribution of states for policy $\mu$, and $y_t$ is defined similarly to Eq. (9).

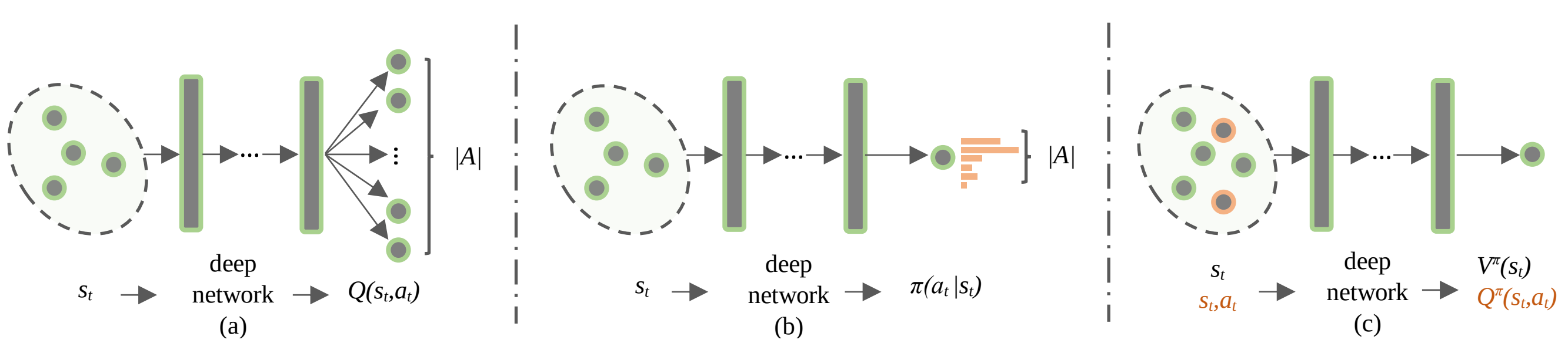

**Fig. 1.** Standard Deep Reinforcement Learning neural network architectures for discrete action spaces. (a) Deep Q-networks approximate the Q-functions for every available action using the state as input. (b) Actor networks approximate the policy distribution over all available actions using the state as input. (c) Critic networks approximate the V-function using the state as input or the Q-function using the state and selected action as input.

Apart from the introduction of deep architectures for action-value functions, two other central features are also integrated in the DQN algorithm that serve training stability and robustness [51]. The first is the *experience replay or replay buffer*, and the second is the use of a separate *target network*. Replay buffer is a memory comprising past experiences that the agent collected by probing $E$, stored in the form of $(s_t, a_t, r_t, s_{t+1})$ tuples. These experiences are used for batch training of the neural network, namely they are sampled uniformly at random by the agent in pre-specified batch sizes at each decision step to approximate the gradient of Eq. (11):

$$\mathbf{g}_{\boldsymbol{\theta}^Q} = \mathbb{E}_{s_t \sim \rho, a_t \sim \mu}\left[\left(Q\left(s_t, a_t \mid \boldsymbol{\theta}^Q\right) - y_t\right)\nabla_{\boldsymbol{\theta}^Q} Q\left(s_t, a_t \mid \boldsymbol{\theta}^Q\right)\right] \tag{12}$$

Except from uniform sampling from past experience, *prioritized sampling* has been shown to work well in many cases, by stochastically directing samples to transition tuples having larger temporal difference values and thus expected to have greater impact on the learning progress [52]. In either case, the sample-based gradient estimate in Eq. (12) is a required input to stochastic gradient descent optimizers, which are essential in deep learning training, for network parameter updates [41]. The target network, $Q^-\left(s_t, a_t \mid \boldsymbol{\theta}^{Q^-}\right)$, duplicates the original network and is used in the calculation of $y_t$:

$$y_t = r\left(s_t, a_t\right) + \gamma \max_{a_{t+1} \in A} Q^-\left(s_{t+1}, a_{t+1} \mid \boldsymbol{\theta}^{Q^-}\right) \tag{13}$$

The target network follows the original network updates in a slower fashion, namely taking values $\boldsymbol{\theta}^{Q^-} = \boldsymbol{\theta}^Q$ with an appropriate delay. DQN with double Q-learning (DDQN) uses both the target and the original networks for computing $y_t$ [53]:

$$y_t = r\left(s_t, a_t\right) + \gamma Q^-\left(s_{t+1}, \arg\max Q\left(s_{t+1}, a_{t+1} \mid \boldsymbol{\theta}^Q\right) \mid \boldsymbol{\theta}^{Q^-}\right) \tag{14}$$

This simple modification of Eq. (13) has been shown to have significant qualities in reducing training instabilities and improving near-optimal solutions by avoiding overoptimistic value estimates. Finally, other advanced variations of DQN include specific *dueling network* architectures that maintain separate layers of *V* and *Q-V* (i.e. advantage) functions before the output layer, facilitating enhanced training in a variety of cases [54]. The specific DQN implementation adopted in this work, for numerical comparison and validation purposes, is presented in detail in Algorithm A1 of Appendix A.

### 3.2. Deep policy gradients

The second major family of DRL algorithms is established on the basis of the policy gradient theorem [34]. Deep policy gradient methods approximate a policy function, $\pi$, with a deep neural network, as previously defined in Eq. (10) and as shown in the *actor* network of Fig. 1(b) for a set of discrete actions, where the output bars denote the probability mass function over all possible actions. Thereby, the policy can be directly updated during training, following the gradient provided by the policy gradient theorem:

$$\mathbf{g}_{\boldsymbol{\theta}^\pi} = \mathbb{E}_{s_{t\geq 0} \sim E, a_{t\geq 0} \sim \pi}\left[\sum_{t\geq 0} \nabla_{\boldsymbol{\theta}^\pi} \log \pi\left(a_t \mid s_t, \boldsymbol{\theta}^\pi\right) Q^\pi\left(s_t, a_t\right)\right] \tag{15}$$

As indicated in [55], Eq. (15) can also take other related forms, and $Q^\pi\left(s_t, a_t\right)$ can be substituted by a class of generic *advantage* functions. The advantage function can be seen as a zero-mean measure, i.e. $\mathbb{E}_{a_t \sim \pi}\left[A^\pi\left(s_t, a_t\right)\right] = 0$, expressing how advantageous an action at each state is, defined as:

$$A^\pi\left(s_t, a_t\right) = Q^\pi\left(s_t, a_t\right) - V^\pi\left(s_t\right) \tag{16}$$

In Eq. (15), computation of the policy gradient requires gradient $\nabla_{\boldsymbol{\theta}^\pi} \log \pi\left(a_t \mid s_t, \boldsymbol{\theta}^\pi\right)$ that is given from the network of Fig. 1(b). Except for gradient $\nabla_{\boldsymbol{\theta}^\pi} \log \pi\left(a_t \mid s_t, \boldsymbol{\theta}^\pi\right)$, the policy gradient requires a complementary estimate related to a certain value. Often, either one of the value or the action-value function is approximated to provide the necessary estimates for Eq. (16) [56, 57], as depicted in the *critic* network of Fig. 1(c). As shown in the figure, in case of an action-value critic, the actions are also required as input to the network (in red). This family of methods is thus referred to as *actor-critic* methods, as the parameters of the policy approximator (actor) are trained with the aid of a value approximator (critic). Other methods to compute the relevant value use Monte Carlo estimates from experience trajectories [58, 59]. Methods relying on function approximations reduce variance but may suffer from increased bias, whereas methods relying on sampling have low bias but high variance. To trade-off bias and variance, some methods in the literature combine both techniques [42, 43, 60, 61]. Along these lines, as proposed in [60], an approximate form of the advantage function in Eq. (16) can be given by:

$$A^\pi\left(s_t, a_t \mid \boldsymbol{\theta}^V\right) \simeq \sum_{i=0}^{k-1} \gamma^i r\left(s_{t+i}, a_{t+i}\right) + \gamma^k V^\pi\left(s_{t+k} \mid \boldsymbol{\theta}^V\right) - V^\pi\left(s_t \mid \boldsymbol{\theta}^V\right) \tag{17}$$

where *k* defines the length of the sampled trajectory the agent actually experienced while probing the environment, and the value function is approximated by a neural network.

Another important distinction in the computation of the policy gradient is also the differentiation between on-policy and off-policy approaches. The gradient in Eq. (15) corresponds to on-policy algorithms. On-policy algorithms are sample inefficient, as opposed to their off-policy counterparts [42, 43], since they require long sampled trajectories, as also indicated by the summation operator in Eq. (15). An efficient method to compute an off-policy gradient estimator of $\mathbf{g}_{\boldsymbol{\theta}^\pi}$, with samples generated by a behavior policy $\mu \neq \pi$, is using importance sampling [42, 43]. In this case, Eq.(15) becomes:

$$\mathbf{g}_{\boldsymbol{\theta}^\pi} = \mathbb{E}_{s_t \sim \rho, a_t \sim \mu}\left[w_t \nabla_{\boldsymbol{\theta}^\pi} \log \pi\left(a_t \mid s_t, \boldsymbol{\theta}^\pi\right) A^\pi\left(s_t, a_t\right)\right] \tag{18}$$

with $w_t = \pi\left(a_t \mid s_t\right) / \mu\left(a_t \mid s_t\right)$. Although this estimator is unbiased,

its variance is high due to the arbitrarily large values $w_t$ can practically take. Truncated importance sampling is a standard approach to cope with high variance in these cases, with $w_t = \min\{c, \pi(a_t \mid s_t)/\mu(a_t \mid s_t)\}$, where $c > 0$ [62].

Actor-critic architectures can also be favorably implemented in problems with continuous actions, which are of particular interest in robotics [63]. Although continuous action spaces can be discretized in order to conform with discrete action approaches, this can be extremely cumbersome for high dimensional actions. Instead, as shown in [56], a deterministic policy gradient also exists in continuous cases, a result that has given rise to deep actor-critic architectures for continuous control tasks [57], which are, however, beyond the scope of this paper and the applications examined herein. These architectures follow the generic schematics of Figs. 1(b) and 1(c) for the actor-critic approximators, with the difference that the actor output layer is not equipped with a discrete probability softmax function and a different policy gradient applies. Alternatively, actor networks following Eqs. (15) and (18) can also be utilized for continuous action distributions, usually based on multivariate Gaussian assumptions with user specified covariance matrices. However, deterministic policy gradients have been shown to have better qualities in continuous action domains [57]. Finally, deep policy gradient methods can be also combined nicely with several features present in DQN training, like experience replay, memory prioritization, dueling architectures, and target networks. All these concepts can significantly enhance algorithmic stability and improve the learning progress.

## 4. Deep Centralized Multi-agent Actor Critic

Maintenance and inspection planning for engineering systems presents great challenges, as already pointed out, in relation to scaling of solution techniques for large problems with multiple components and component states and actions. In the most comprehensive and detailed control cases, the problem is fully formulated in spaces that scale exponentially with the number of components, making solutions practically intractable by conventional planning and learning algorithms, without resorting to simplified, less accurate modeling approaches that reduce complexity. DRL provides a valuable framework for dealing with state space complexity, enabling advanced management solutions in high-dimensional domains. All previously presented DRL approaches, from DQNs to deep policy gradients, provide exceptional parametrization capabilities for huge state spaces. However, this is not the case for similarly large discrete action spaces, which are also present in multi-component systems. Note for example that the output layer of DQNs consists of a Q-value for each available action, thus defining an output space of $|A|$ dimensions. Apparently, this network structure cannot support a sizable number of distinct actions, hence it is only appropriate for systems with small action spaces. Similarly, deep policy gradient architectures include actor networks that output a probability distribution over all possible system actions.

In [64] a so-called Wolpertinger architecture has been proposed for large discrete action spaces, combining a deep deterministic policy gradient network [57] with a nearest neighbor action-space reduction. In this approach, however, the implemented nearest neighbor layer introduces discontinuities that interrupt the differentiability of the network, potentially leading to severe training instabilities due to improper backpropagation of gradients. In [65] dueling DQNs are modified to incorporate action branching architectures, in an effort to alleviate the complexity of discrete actions emerging from the discretization of continuous action spaces. This approach essentially assumes a decomposition of Q-values for different continuous dimensions, allowing separate value maximizations for each dimension.

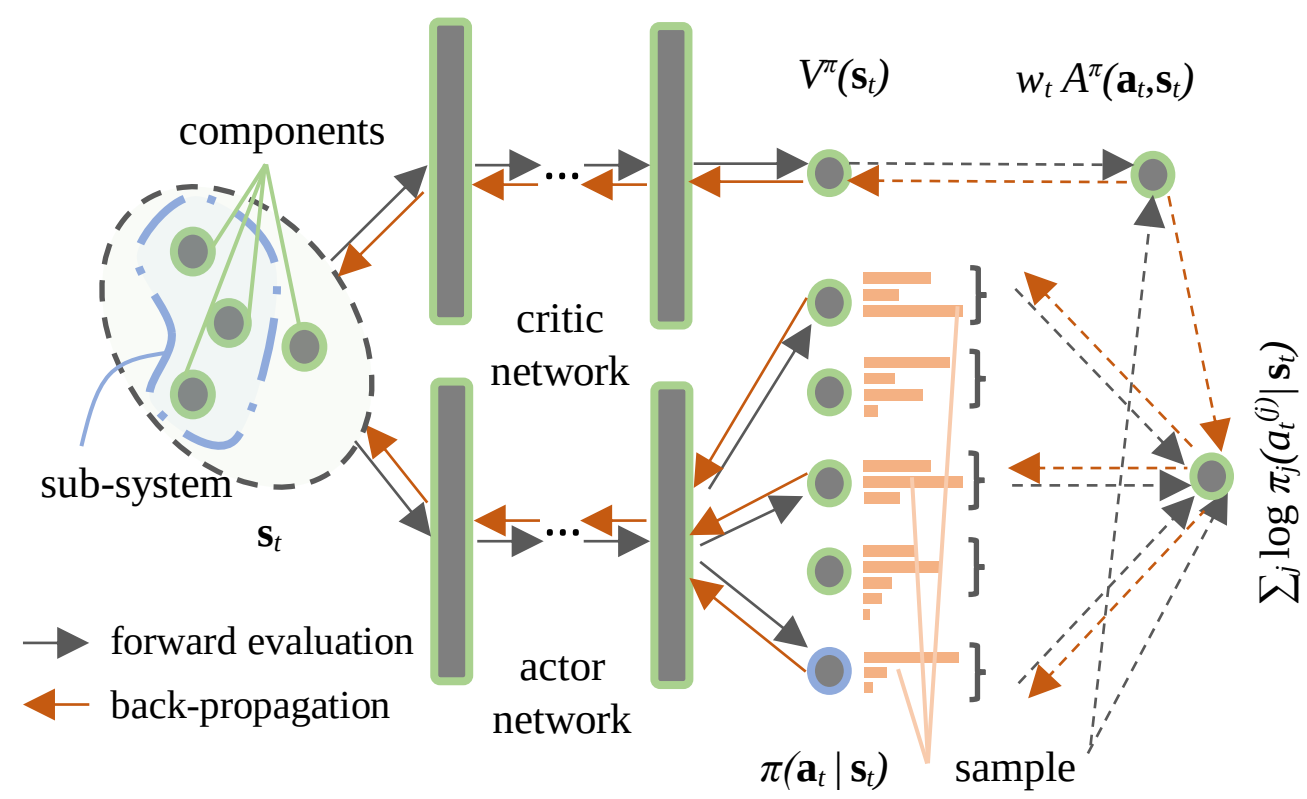

**Fig. 2.** Deep Centralized Multi-agent Actor Critic (DCMAC) architecture. Forward pass for function evaluations (black links) and weighted advantage back-propagation for training (red links). Dashed lines represent operations and dependencies that do not involve deep network parameters.

In this work, to concurrently tackle the state and action scalability issues, we develop and introduce a new approach, the *Deep Centralized Multi-agent Actor Critic* (DCMAC), within the premises of off-policy actor-critic DRL with experience replay. This approach takes advantage of the parametrization capabilities of deep policy gradient methods in large state spaces, and further extends them to large action spaces that have numerous distinct action categories. Actions adhere to a specific probability structure that can drastically alleviate the complexity related to the output layer. Specifically, DCMAC naturally assumes that actions on system components, as well as sub-system actions (compound actions with effects on greater parts of the system), are conditionally independent of each other, given the state of the entire system:

$$\pi(\mathbf{a}_t \mid \mathbf{s}_t) = \prod_{i=1}^{n} \pi_i\left(a_t^{(i)} \mid \mathbf{s}_t\right) \tag{19}$$

where $n$ is the number of *control units*, which can include components and greater sub-system parts on which individualized actions apply, $\mathbf{a}_t = \{a_t^{(i)}\}_{i=1}^{n}$, and $\mathbf{s}_t = \{s_t^{(i)}\}_{i=1}^{m}$, with $m$ denoting the number of system components. A component is defined as a minimum structural entity of the system for which a separate state variable exists. In the schematic of the DCMAC in Fig. 2, a system with $m$=4 components is shown, with their states given as input to the suggested actor and critic networks. Accordingly, there are 4 sets of available actions, one for each component, at the actor output, together with 1 set of available actions describing sub-system decisions that pertain to a group of components. As such, the total number of control units in the figure is $n$=5, and the related output bars denote the probability mass functions over all possible specific actions for each control unit. This actor architecture technically means then that every control unit operates as an autonomous agent that utilizes centralized system-state information to decide about its

**Algorithm 1** Deep Centralized Multi-agent Actor Critic (DCMAC).

Initialize replay buffer
Initialize actor and critic network weights $\boldsymbol{\theta}^\pi$ , $\boldsymbol{\theta}^V$
**for** *episode* = 1,*M* **do**
  **for** *t*=1,*T* **do**
    Select action $\mathbf{a}_t$ at random according to exploration noise
    Otherwise select action $\mathbf{a}_t \sim \boldsymbol{\mu}_t = \pi\left(\cdot \mid \hat{\mathbf{b}}_t, \boldsymbol{\theta}^\pi\right)$
    Collect reward $r\left(\hat{\mathbf{b}}_t, \mathbf{a}_t\right)$ sampling $\hat{\mathbf{b}}_t$
    Observe $o_{t+1}^{(l)} \sim p\left(o_{t+1}^{(l)} \mid \mathbf{b}_t^{(l)}, \mathbf{a}_t\right)$ for $l = 1,2,\dots,m$
    Compute beliefs $\mathbf{b}_{t+1}^{(l)}$ for $l = 1,2,\dots,m$
    $$b^{(l)}\left(s_{t+1}^{(l)}\right) = \frac{p\left(o_{t+1}^{(l)} \mid s_{t+1}^{(l)}, \mathbf{a}_t\right)}{p\left(o_{t+1}^{(l)} \mid \mathbf{b}_t^{(l)}, \mathbf{a}_t\right)} \sum_{s_t^{(l)} \in S^{(l)}} p\left(s_{t+1}^{(l)} \mid s_t^{(l)}, \mathbf{a}_t\right) b^{(l)}\left(s_t^{(l)}\right)$$
    Store experience $\left(\hat{\mathbf{b}}_t, \mathbf{a}_t, \boldsymbol{\mu}_t, r\left(\hat{\mathbf{b}}_t, \mathbf{a}_t\right), \hat{\mathbf{b}}_{t+1}\right)$ to replay buffer
    Sample batch of $\left(\hat{\mathbf{b}}_i, \mathbf{a}_i, \boldsymbol{\mu}_i, r\left(\hat{\mathbf{b}}_i, \mathbf{a}_i\right), \hat{\mathbf{b}}_{i+1}\right)$ from replay buffer
    If $\hat{\mathbf{b}}_{i+1}$ is terminal state $A_i = r\left(\hat{\mathbf{b}}_i, \mathbf{a}_i\right) - V\left(\hat{\mathbf{b}}_i \mid \boldsymbol{\theta}^V\right)$
    Otherwise $A_i = r\left(\hat{\mathbf{b}}_i, \mathbf{a}_i\right) + \gamma V\left(\hat{\mathbf{b}}_{i+1} \mid \boldsymbol{\theta}^V\right) - V\left(\hat{\mathbf{b}}_i \mid \boldsymbol{\theta}^V\right)$
    Update actor parameters $\boldsymbol{\theta}^\pi$ according to gradient:
    $$\mathbf{g}_{\boldsymbol{\theta}^\pi} \simeq \sum_i w_i \left( \sum_{j=1}^{n} \nabla_{\boldsymbol{\theta}^\pi} \log \pi_j \left(a_t^{(j)} \mid \hat{\mathbf{b}}_i, \boldsymbol{\theta}^\pi\right)\right) A_i$$
    Update critic parameters $\boldsymbol{\theta}^V$ according to gradient:
    $$\mathbf{g}_{\boldsymbol{\theta}^V} \simeq \sum_i w_i \nabla_{\boldsymbol{\theta}^V} V^\pi\left(\hat{\mathbf{b}}_i \mid \boldsymbol{\theta}^V\right) A_i$$
  **end for**
**end for**

actions. Relevant deep network architectures for determining cooperative/competitive multi-agent strategies have been developed using deterministic policy gradients and off-policy learning with Q-function critics [66], or by assuming identical actor networks for all agents and on-policy learning with Q-function critics [67]. The benefit of our suggested representation becomes even clearer when we substitute Eq. (19) in Eq. (18), to obtain the policy gradient:

$$\mathbf{g}_{\boldsymbol{\theta}^\pi} = \mathbb{E}_{\mathbf{s}_t \sim \boldsymbol{\rho}, \mathbf{a}_t \sim \boldsymbol{\mu}} \left[ w_t \left( \sum_{i=1}^{n} \nabla_{\boldsymbol{\theta}^\pi} \log \pi_i \left(a_t^{(i)} \mid \mathbf{s}_t, \boldsymbol{\theta}^\pi\right)\right) A^\pi\left(\mathbf{s}_t, \mathbf{a}_t\right)\right] \tag{20}$$

where $\boldsymbol{\mu}$ is now a $n$-dimensional vector of agents' behavior policies, and $\boldsymbol{\rho}$ is the $m$-dimensional limiting state distribution under these policies. Eq. (20) implies a particularly convenient representation of the actor. As also shown for the system of Fig. 2, the 5 control units are equipped with 3, 4, 3, 5, and 3 actions to choose from, respectively, from top to bottom. Hence, this system with a total of 540 possible actions is scaled down with the DCMAC actor to an output action space of 18 dimensions, without any loss of generality. In Eq. (20) it is also observed, similarly to Fig. 2, that the multiple agents are not described by independent networks but are supported by a centralized actor with shared parameters, $\boldsymbol{\theta}^\pi$. As such, not only does each agent know the states of all other agents as a result of taking the entire system state, $\mathbf{s}_t$, as input, but it is also implicitly influenced by their actions through $\boldsymbol{\theta}^\pi$.

As indicated in Eq. (20) and Fig. 2, the importance sampling weighted advantage needs to be determined and back-propagated for the computation of gradients. The single-valued output of the critic, together with sampled actions for each component are used for the evaluation of the advantage function, similarly to Eq. (17). However, an off-policy implementation of the advantage function in Eq. (17) would require a product of $k$ importance sampling weights, as $k$ independent transition samples are involved. This fact, along with the required product of weights resulting from the factorized representation of the actor output in Eq. (19), could increase the variance of the estimator significantly. Thus, the utilized advantage function is computed by Eq. (17) for $k$=1 here, essentially following the temporal difference:

$$A^\pi\left(\mathbf{s}_t, \mathbf{a}_t \mid \boldsymbol{\theta}^V\right) \simeq r\left(\mathbf{s}_t, \mathbf{a}_t\right) + \gamma V^\pi\left(\mathbf{s}_{t+1} \mid \boldsymbol{\theta}^V\right) - V^\pi\left(\mathbf{s}_t \mid \boldsymbol{\theta}^V\right) \tag{21}$$

Apart from the actor network, the critic network is also centralized, as implied by Eq. (21) and shown in Fig. 2. The critic approximates the value function over the entire system space, thus providing a global measure for the DCMAC policy updates. The critic is updated through the mean squared error, similarly to Eq. (11):

$$L_V\left(\boldsymbol{\theta}^V\right) = \mathbb{E}_{\mathbf{s}_t \sim \boldsymbol{\rho}, \mathbf{a}_t \sim \boldsymbol{\mu}} \left[ w_t \left( r\left(\mathbf{s}_t, \mathbf{a}_t\right) + \gamma V^\pi\left(\mathbf{s}_{t+1} \mid \boldsymbol{\theta}^Q\right) - V^\pi\left(\mathbf{s}_t \mid \boldsymbol{\theta}^Q\right)\right)^2 \right] \tag{22}$$

As shown in Fig. 2, the respective gradient is computed by back-propagating the weighted advantage function through the critic network:

$$\mathbf{g}_{\boldsymbol{\theta}^V} = \mathbb{E}_{\mathbf{s}_t \sim \boldsymbol{\rho}, \mathbf{a}_t \sim \boldsymbol{\mu}} \left[ w_t \nabla_{\boldsymbol{\theta}^V} V^\pi\left(\mathbf{s}_t \mid \boldsymbol{\theta}^V\right) A^\pi\left(\mathbf{s}_t, \mathbf{a}_t \mid \boldsymbol{\theta}^V\right)\right] \tag{23}$$

As indicated by Eqs. (20) and (23) DCMAC operates off-policy. This is driven by the fact that off-policy algorithms are more sample efficient than their on-policy counterparts, as previously underlined. This attribute can be critical in large engineering systems control, as samples are often drawn from computationally expensive nonlinear and/or dynamic structural models through demanding numerical simulations, e.g. in [68, 69]. In addition, as in most standard DRL approaches, experience replay is utilized here for more efficient training.

As already discussed, observations in structural and engineering systems are in general unlikely to be able to reveal the actual state of the system with certainty, which makes the problem more adequately described by POMDPs. Following the concept of belief MDPs, as suggested by Eqs. (6) and (7), any DRL network can be implemented as a belief DRL network, if a model for the component transition and observation matrices, $p\left(s_{t+1}^{(i)} \mid s_t^{(i)}, \mathbf{a}_t\right)$ and $p\left(o_t^{(i)} \mid s_t^{(i)}, \mathbf{a}_t\right)$, for all $i = 1,2,\dots,m$, is now known. This is a valid assumption for observations in engineering systems, since inspection methods and monitoring instruments are in general accurate up to known levels of precision. In this case, all Eqs. (11)-(23) still hold, except states $\mathbf{s}_t$ are substituted by beliefs $\hat{\mathbf{b}}_t = \left\{\mathbf{b}_t^{(i)}\right\}_{i=1}^{m}$, which are continuously updated over the life-cycle, based on the selected actions and observations. A similar concept has been applied for DQNs in [70], where it is shown in benchmark small POMDP applications that belief DQNs are able to provide adequate nonlinear approximations of the piece-wise linear value function of Eq. (8).

In Algorithm 1, the pertinent algorithmic steps for the off-policy learning algorithm developed to implement DCMAC is shown. Algorithm 1 describes the procedure corresponding to the belief MDPs. In the case of MDPs, where the environment is completely observable by the agents, i.e. $o_t^{(i)} = s_t^{(i)}$, component belief vectors, $\mathbf{b}_t^{(i)}$, are substituted by the exact component states, $s_t^{(i)}$, and the Bayesian updates described in Eq. (6) are omitted. It should be also noted that under complete observability, i.e. in MDP settings, no model for component observations and state transitions needs to be known, whereas for POMDPs, explicit models are necessary for belief updates, but only for the individual components and not for the entire system.

## 5. Numerical experiments

To explore the applicability of DRL in engineering system problems and to evaluate the proposed framework, we run numerical experiments in reference to life-cycle cost minimization for three different systems that cover a great span of learning settings, regarding the number of system states, structural dependencies between components, non-stationarity of transitions, number of available actions per control unit, accuracy level of observations, and availability of inspection choices. More specifically, the examined finite horizon examples involve: (i) a simple stationary parallel-series MDP system, (ii) a non-stationary system with k-out-of-n modes in both MPD and POMDP environments, and (iii) a bridge truss system subject to non-stationary corrosion, simulated through an actual nonlinear structural model, in a POMDP environment, including optional selection of observation actions.

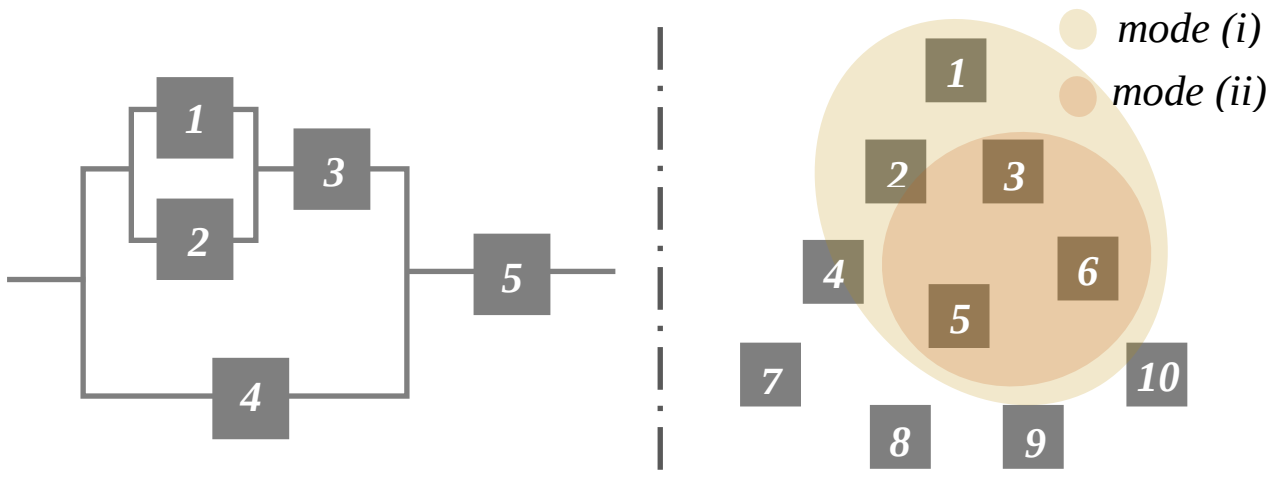

**Fig. 3.** (a) System I: Parallel-series system. (b) System II: System with k-out-of-n modes. Contours indicate an example of system modes activation. Lighter indicates components being at greater than severe damage state, and darker indicates components being at failure damage state.

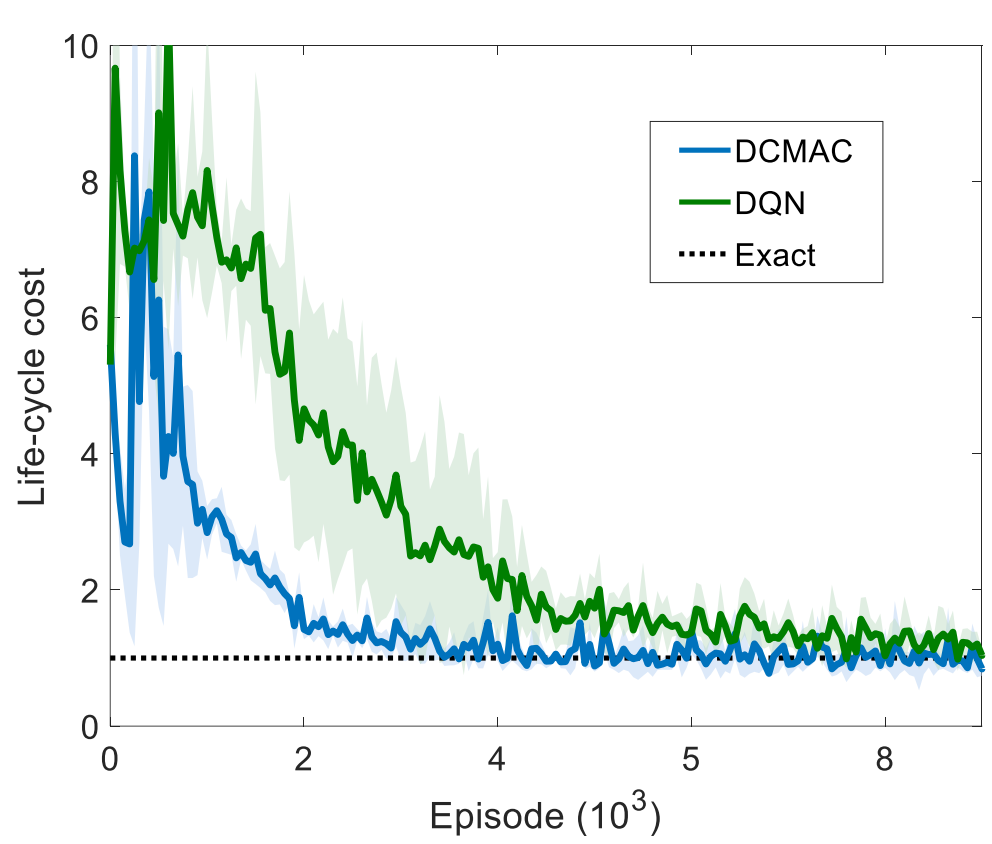

**Fig. 4.** Life-cycle cost during training for parallel-series system (System I), with 95% confidence intervals, using DCMAC and DQN approaches, normalized with respect to the exact policy.

### 5.1. System I: Stationary parallel-series system

This simple parallel-series system, shown in Fig. 3(a), consists of 5 components. There are 4 states per component, indicating damage states of increasing severity, i.e. *no damage* (state 1), *minor damage* (state 2), *severe damage* (state 3), and *failure* (state 4). Interstate Markovian dynamics are defined by unique stationary transition matrices for each component. Matrices are triangular, culminating in the final absorbing failure state, to emulate a standard deteriorating environment. More details on component state transitions, as well as related costs, can be found in Appendix B.

System failure is defined by the pertinent combinations of failed components, as suggested by the system structure in Fig. 3(a). When the system fails, component state costs are penalized by a factor of 24. Each component defines a separate control unit, having 2 actions available, i.e. *do nothing* (action 1) and *replace* (action 2). Do nothing has no effect on the component state, whereas replace sends the component to transition from the no damage state. Actions are taken at every step, for a planning horizon, or else episode, of 50 steps, e.g. years, and a discount factor of $\gamma = 0.99$ is used. High discount factor values extend the effective horizon of important decisions, thus making the problem more challenging, as more future actions matter. Overall, there are 1,024 states, and 32 actions for this system and an exact MDP solution is tractable through value iteration. Thereby, this basic experimental setup serves for evaluating the accuracy and quality of DRL solutions. In this problem, we use both DQN and DCMAC to obtain near-optimal policies. DQN is applicable here as well due to the small number of system actions, and it is also implemented for verification purposes. Details about the network specifications can be found in Appendix C.

In Fig. 4, the mean life-cycle cost collected during training from 5 runs is presented, together with its corresponding 95% confidence intervals, under Gaussian assumptions. As can be observed, DCMAC exhibits a better anytime performance and has faster convergence compared to DQN, however both algorithms eventually converge to the exact solution with less than 5% error. In Fig. 5, an indicative policy realization for the best run of each algorithm is illustrated, for all system components. In this particular realization, the DCMAC policy becomes identical to the exact policy (not shown in figure for clarity) after about 10 thousand episodes. DQN partly diverges from this solution, as seen for components 1 and 3. This is a general pattern since, as thoroughly checked through multiple policy realizations, action agreement between exact policy and the two algorithms is about 99% and 96%, for DCMAC and DQN, respectively.

### 5.2. System II: Non-stationary system with k-out-of-n modes

The examined system in this example has 10 components, with each one having 4 states. For the characterization of damage, the same state definitions, as in System I, are considered. Transitions are now non-stationary, depending also on the deterioration rate of each component. Appendix B provides all the additional necessary details that completely describe the modeled environment in this example. In accordance with the performance of a great variety of engineering systems, state combinations of individual components are considered to trigger system damage modes of different severity, following a k-out-of-n activation function. More specifically, two discrete system damage modes are considered; mode (i) is activated when the number of components in at least state 3 is equal or exceeds 50% of the total number of system components, and mode (ii) when the number of components at state 4 is equal or exceeds 30%, as also schematically depicted in Fig. 3(b). Component-wise direct costs continue to follow the damage state cost functions as in the previous example, however they are penalized by a factor of 2.0 and 12.0 when modes (i) and (ii) are activated, respectively. When both modes are active the penalty factor is 24.0. Episode length is 50 years and the discount factor is again $\gamma = 0.99$.

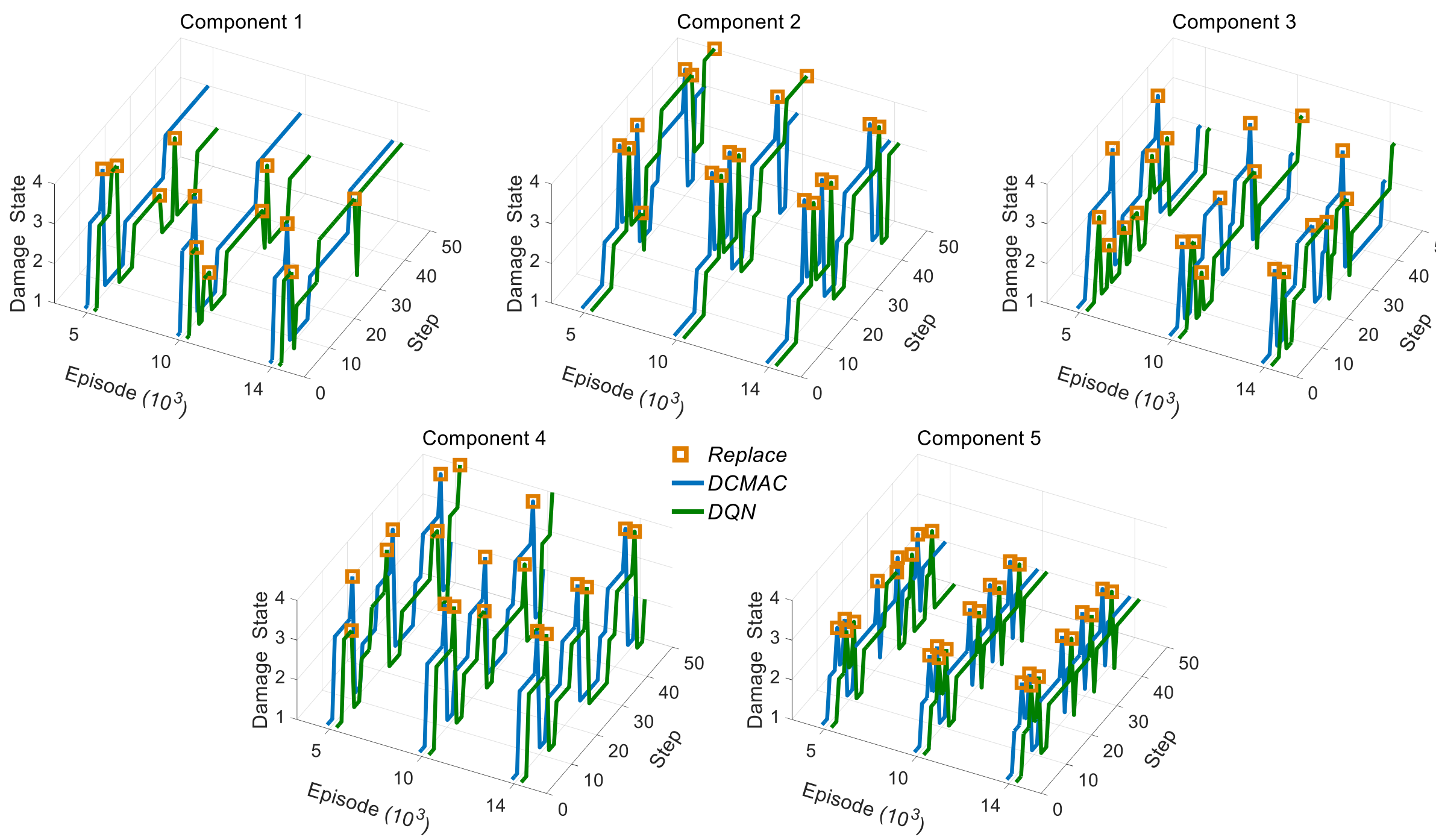

**Fig. 5.** Policy realizations for all System I components at three different training episodes, based on DCMAC and DQN solutions. All component policies converge to the exact solution for both methods after 10 thousand episodes, expect for components 1 and 3 for the DQN solution.

There are 4 available actions per component, i.e. *do nothing* (action 1), *minor repair* (action 2), *major repair* (action 3), and *replace* (action 4). Do nothing leaves component state and deterioration rate unchanged, minor repair only reduces component state by 1, at that decision step and before the next transition, with a success probability of 0.95, and major repair has the same effect on component states as the minor repair, but in addition reduces the component deterioration rate by 5 steps. Replace sends components to an intact condition, namely back to state 1 and the initial deterioration rate, again at that decision step and before the next transition. As discussed in [18] such actions can effectively describe maintenance decisions for certain types of concrete structures, among others. As deterioration rate essentially reflects the effective age of the components, major repairs not only improve component states but also suspend their aging process, which, as time passes, prompts transitions to more severe states in the uncontrolled case due to the non-stationarity of the environment.

System II is examined both under complete and partial observability. In the former case the problem is formulated as a MDP, which means that the agent observes the exact state of the system at each decision step. In the latter case, observations do not reveal the exact state of the system with certainty, so the agent forms a belief about its state, computed by Eq. (6). Four different observation cases are considered, reflecting the accuracy of the inspection instruments, with accuracy levels $p = 1.0, 0.9, 0.8, 0.7$. The value of $p$ indicates the probability of observing the correct component state, thus $p = 1.0$ defines the MDP case. More details are provided in Appendix B. For the POMDP cases, with $p = 0.9, 0.8, 0.7$, partial observability only applies to the component damage states, whereas the deterioration rate is considered to be known, thus overall the problem can be also seen as a MOMDP.

For this system an exact solution is not available due to the high dimensionality of the state and action spaces. As a result of different component damage state configurations combined with different possible deterioration rates, the total number of states is greater than $10^{23}$, whereas the total number of available system actions is equal to $10^{6}$. This total number of states and actions is indicative of how the curse of dimensionality can impede comprehensive maintenance and inspection solutions in multi-component systems and underscores the imperative strengths that DRL with DCMAC provides. The implemented deep network specifications and details are discussed in Appendix C. DQN is not applied in this case due to the large number of actions, which forces the number of network parameters to explode and, thus, hard to train. Therefore, to obtain relative policy measures in order to assess and evaluate the DCMAC solutions, we formulate and evaluate 4 straightforward baseline policies, optimized under complete observability, indicative of standard engineering practice in maintenance planning, incorporating condition-based maintenance (CBM) and time-condition-based maintenance (TCBM) considerations. The first baseline, CBM-I, is a simple condition-based replacement policy. The decision at each step is whether to replace a component or not, merely based on its state of damage. According to the optimal replacement strategy for this policy, components reaching or exceeding severe damage (state 3) are replaced. The second baseline, CBM-II, takes also into account the possibility of repair actions based on the damage state of the components, and more specifically of action 2 which has only damage corrective effects. Along these lines, the optimal CBM-II baseline chooses actions based on the observed current damage state, assigning do nothing action to no damage, minor repair to minor and severe damage, and replace to failure. The last two baselines, TCBM-I and TCBM-II respectively, are the same as their CBM counterparts with regard to state-action pairs, however they also account for major repairs or replacements when a component deterioration rate reaches or exceeds certain levels. A 5-year limit for the deterioration rate corresponds to the optimized TCBM-I, which suggests a major repair every time this limit is reached at the minor damage state. Similarly for the optimum TCBM-II, the deterioration rate limit is 5 years, suggesting major repair actions at states of minor and severe damage.

**Table 1**
System II life-cycle cost estimates and 95% confidence intervals under complete and partial observability, for inspections of accuracy $p$=1.0, 0.9, 0.8, 0.7 (normalization with respect to MDP DCMAC solution).

| Observability | DCMAC | CBM-I | CBM-II | TCBM-I | TCBM-II |
|---|---|---|---|---|---|
| p = 1.0 | 1.0000 ± 0.0077 | 1.4350 ± 0.0093 | 1.2239 ± 0.0062 | 1.4207 ± 0.0082 | 1.1374 ± 0.0054 |
| p = 0.9 | 1.0442 ± 0.0074 | 1.5291 ± 0.0092 | 1.3725 ± 0.0061 | 1.5506 ± 0.0081 | 1.3118 ± 0.0054 |
| p = 0.8 | 1.0790 ± 0.0076 | 1.6069 ± 0.0086 | 1.5164 ± 0.0059 | 1.6560 ± 0.0081 | 1.4789 ± 0.0057 |
| p = 0.7 | 1.1036 ± 0.0081 | 1.6809 ± 0.0088 | 1.6647 ± 0.0059 | 1.7518 ± 0.0086 | 1.6617 ± 0.0056 |

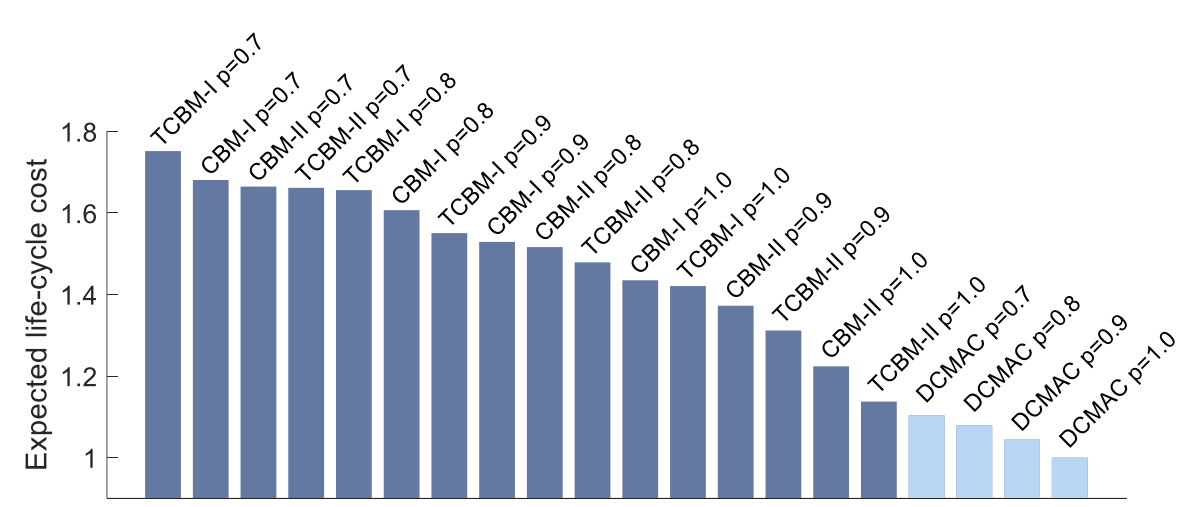

**Fig. 6.** Expected life-cycle cost estimates for System II, for best DCMAC solution, all baseline policies, and different observability accuracies.

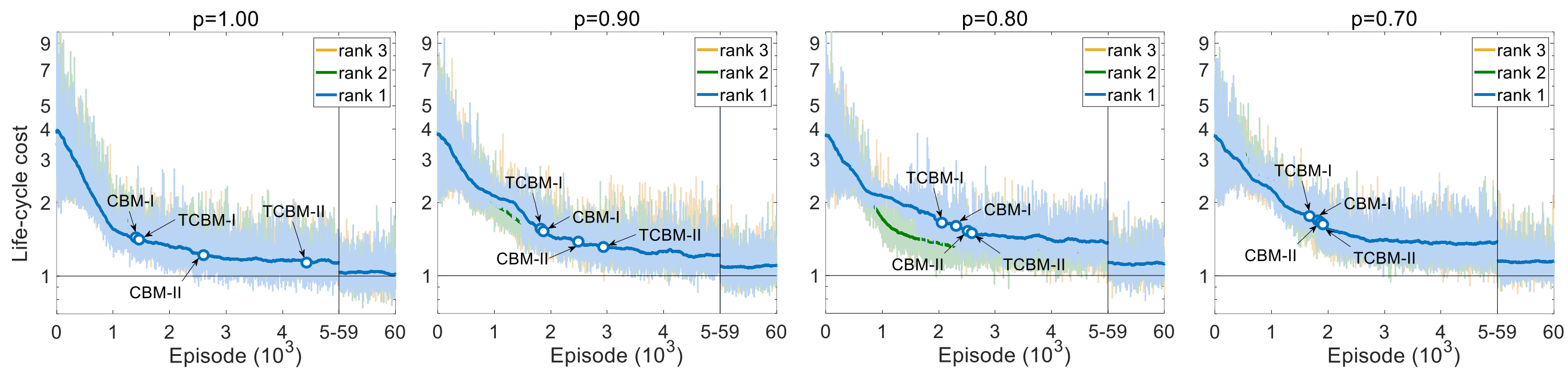

**Fig. 7.** Expected life-cycle cost estimates during training for System II, for top three DCMAC solutions. Baseline policy estimates are indicated on the mean curve of the best solution.

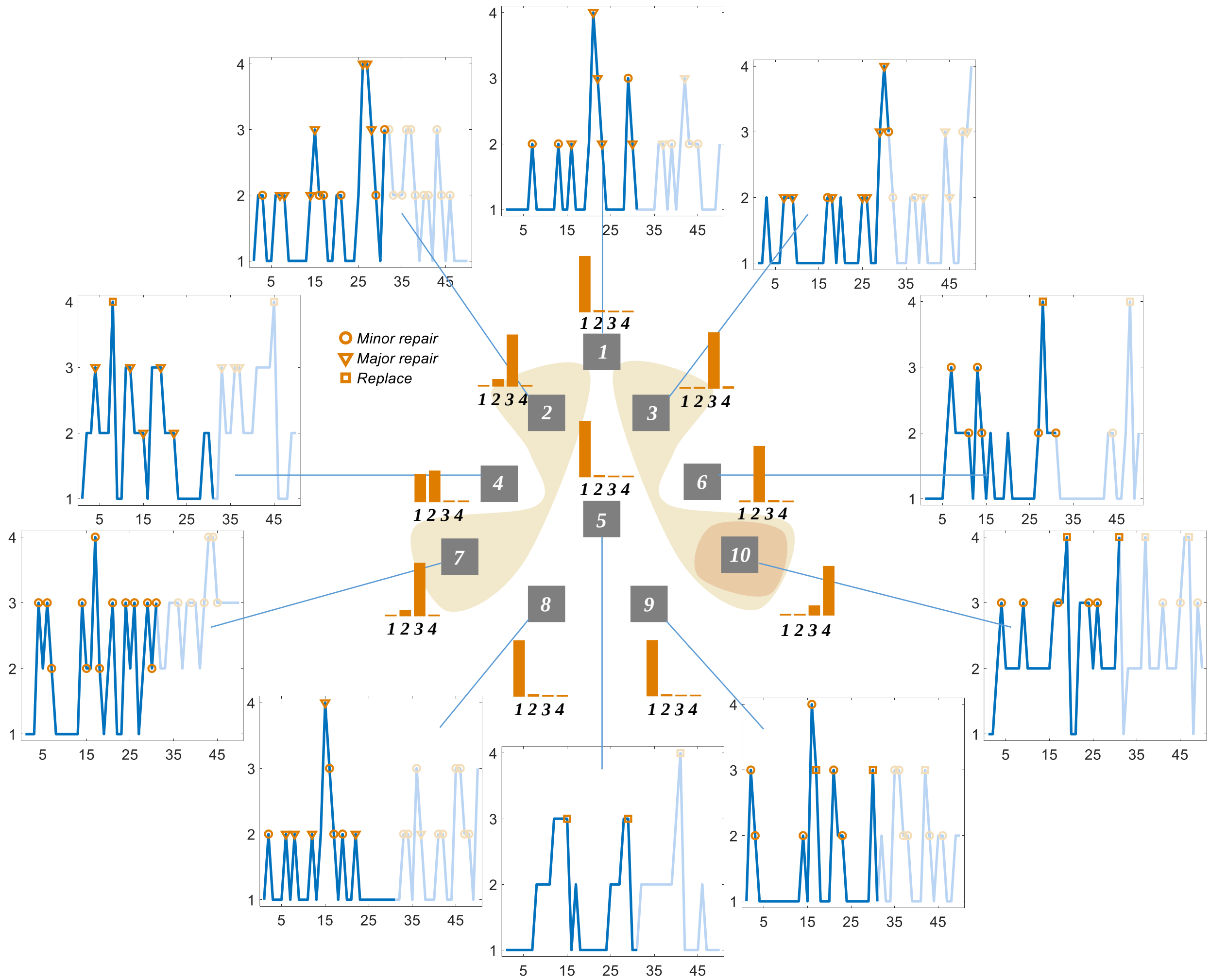

**Fig. 8.** Policy realization of System II with snapshot of damage state beliefs for all components, for observation accuracy $p$=0.90, at time step $t$=31 years. Contours depict components with observed damage states 3 and 4. Plots show the evolution of observed damage state over the planning horizon of 50 years, together with the selected maintenance actions at different time steps.

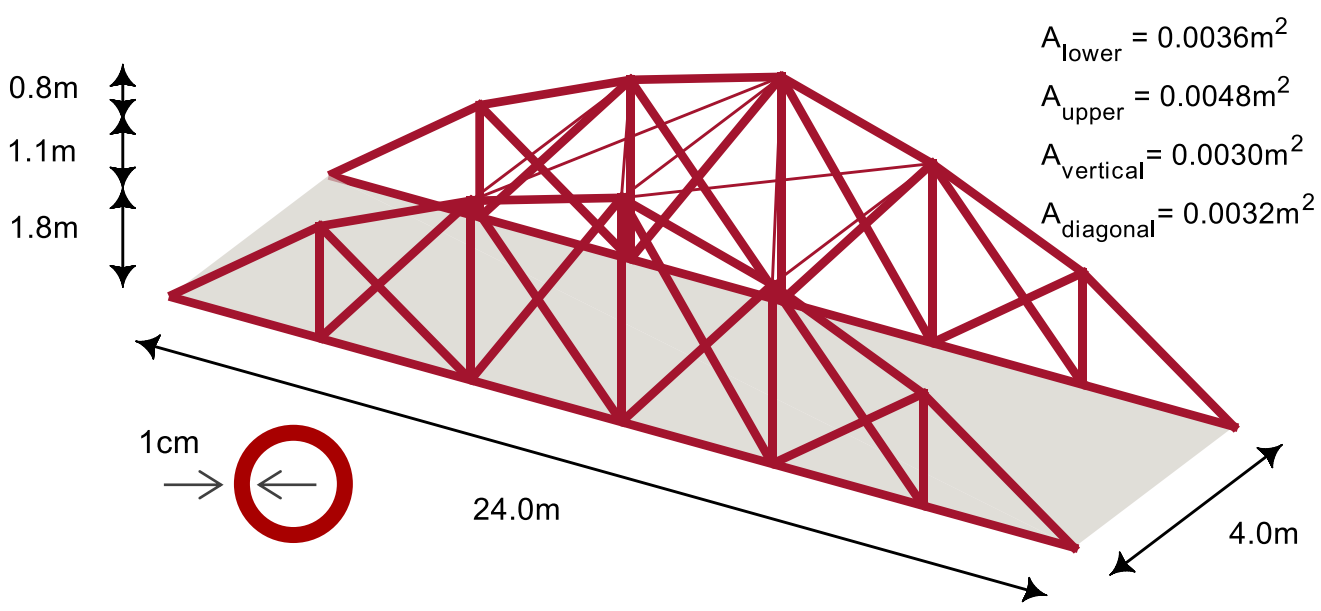

**Fig. 9.** Geometrical properties and cross-section areas of steel truss bridge structure (System III).

The results of the best DCMAC run and the abovementioned baselines are presented in Table 1, where values are calculated through $10^3$ Monte Carlo 50-step policy realizations and are normalized with respect to the results of the MDP case, namely for $p=1.0$. The baselines are optimized, as mentioned, for the case of complete observations, whereas their implementation in the partially observable domains are based on the observed damage state at each decision step. As seen in the table, for the various observability levels examined, DCMAC discovers substantially better policies.

The competence of DCMAC policies is also indicated by the fact that they are also superior even when the baselines operate under better observability conditions, as can be seen in Fig. 6, whereas as depicted in Fig. 7, this is even accomplished before 3.0 thousand episodes, except for TCBM-II under perfect observability, which is surpassed after 4.4 thousand episodes. Reflecting on the DCMAC results, it can be seen that observability is critical for life-cycle cost problems, as expected. That is, more accurate inspections lead to better system state determinations, and thus improved decisions in terms of total expected cost in the long run. The general pattern of increased life-cycle cost as observations become less informative is also apparent in the baseline policies, as shown both in Table 1 and Fig. 6. The difference of total expected costs for policies with inspections characterized by different precision levels quantifies the notion of the Value of Information (VoI) or the value of more precise information, which can have thorough implications regarding proper selection of monitoring systems.

In Fig. 8, a policy realization is depicted for the case of $p=0.90$. A snapshot of the system damage state belief is captured at $t$=31 years, illustrated in the form of probability bars over possible damage states of each component. The plots for each component display the observed damage state evolution, in addition to the actions taken throughout the control history of the 50 years horizon. The future states and actions are shown with lighter color as well. An interesting remark about this snapshot is that the system environment at this time, as the used contours also demonstrate, is at the brink of activation of modes (i) and (ii), the combination of which practically suggests extensive losses at the system level. As such, control actions are taken for all components with observed states greater than 1, which would not be the case otherwise. Another general pattern in the belief evolution is that recent replace actions or consecutive repairs, except for drastically reducing damage, also result in more certainty about the actual damage state of a component, as indicatively shown for components 1, 3, 5, 6 and 9. Conversely, when no control action is taken for a prolonged time, damage state beliefs are more likely to be less informative as, for example, can be observed for component 4. Overall, as observed, DCMAC provides a very detailed and complex life-cycle policy for all individual system components, without needing any unnecessary a priori policy constraints to enable solutions.

**Table 2**
System III baseline policy specifications and life-cycle cost estimates with 95% confidence intervals (normalization with respect to DCMAC solution).

| Policy Name | Inspection period (years) | Major repair threshold | Replace threshold | Life-cycle cost |
|---|---|---|---|---|
| CBM-I | 2 | 10.0% | 15.0% | 1.2227 ± 0.0026 |
| CBM-II | 5 | 10.0% | 12.5% | 1.0772 ± 0.0025 |
| CBM-III | 10 | 10.0% | 12.5% | 1.0576 ± 0.0023 |
| CBM-IV | 15 | 10.0% | 12.5% | 1.0633 ± 0.0023 |

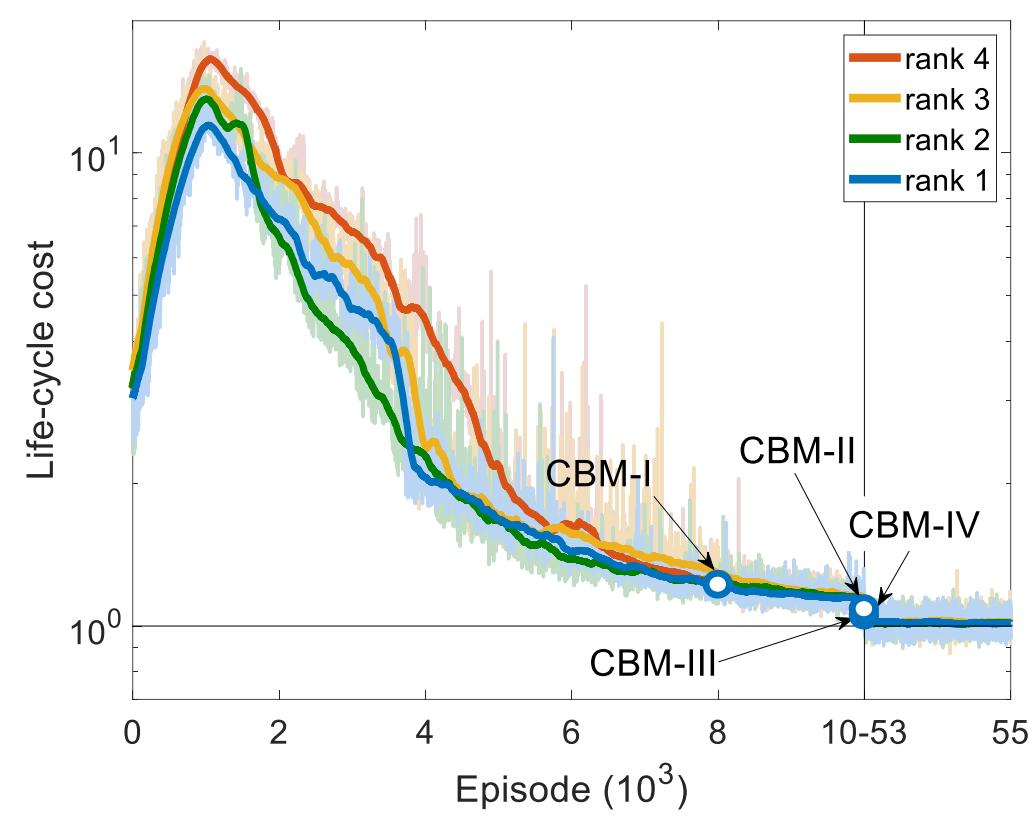

**Fig. 10.** System III expected life-cycle cost during training for top four DCMAC solutions and baseline policies. Baseline policy estimates are indicated on the mean curve of the best solution.

## 5.3. System III: Non-stationary truss bridge structure

In this example, DCMAC is implemented in a structural setting. System III is a steel bridge structure, consisting of truss members with hollow circular cross-sections of different areas, as illustrated in Fig. 9. The material yield stress is 355MPa, following a Bouc-Wen-type 2% linear kinematic hardening, as discussed in [71], whereas buckling of the compressed members, along with other geometrical nonlinearities, is ignored here. The structure is imposed to a uniformly distributed vertical load, applied at the deck level, which is considered to follow a normal distribution with a mean value of 16.25kN/m$^2$ and a 10% coefficient of variation. The two truss sub-structures composing the entire structural system are considered to be structurally independent, thus being subject to the same vertical loads at every time step, and their truss members are separately maintained. As a result of this structural and loading symmetry, the total number of components involved in the analysis is 25. The structure is considered to operate in a corrosive environment, which inflicts section losses to all components. Section losses are independent for each component, modeled as non-stationary gamma processes. By proper discretization, the underlying continuous space of section losses is divided in 25 discrete damage states for each component. Details about the continuous stochastic processes evolution, formation of the damage state transition matrices and computation of pertinent component costs are included in Appendix B. There exist four system modes in this example that indirectly define component configurations with global effects. These are defined by the normalized displacement of the middle lower node of the truss, $u$, with respect to the ultimate value of this displacement at yielding, $u_y$, for the intact condition without corrosion. For thresholds $u/u_y = 0.60, 0.75, 1.00$, direct component state costs are penalized by 2.0, 6.0, 24.0, respectively. The planning horizon of the problem is 70 years, with $\gamma = 0.99$.

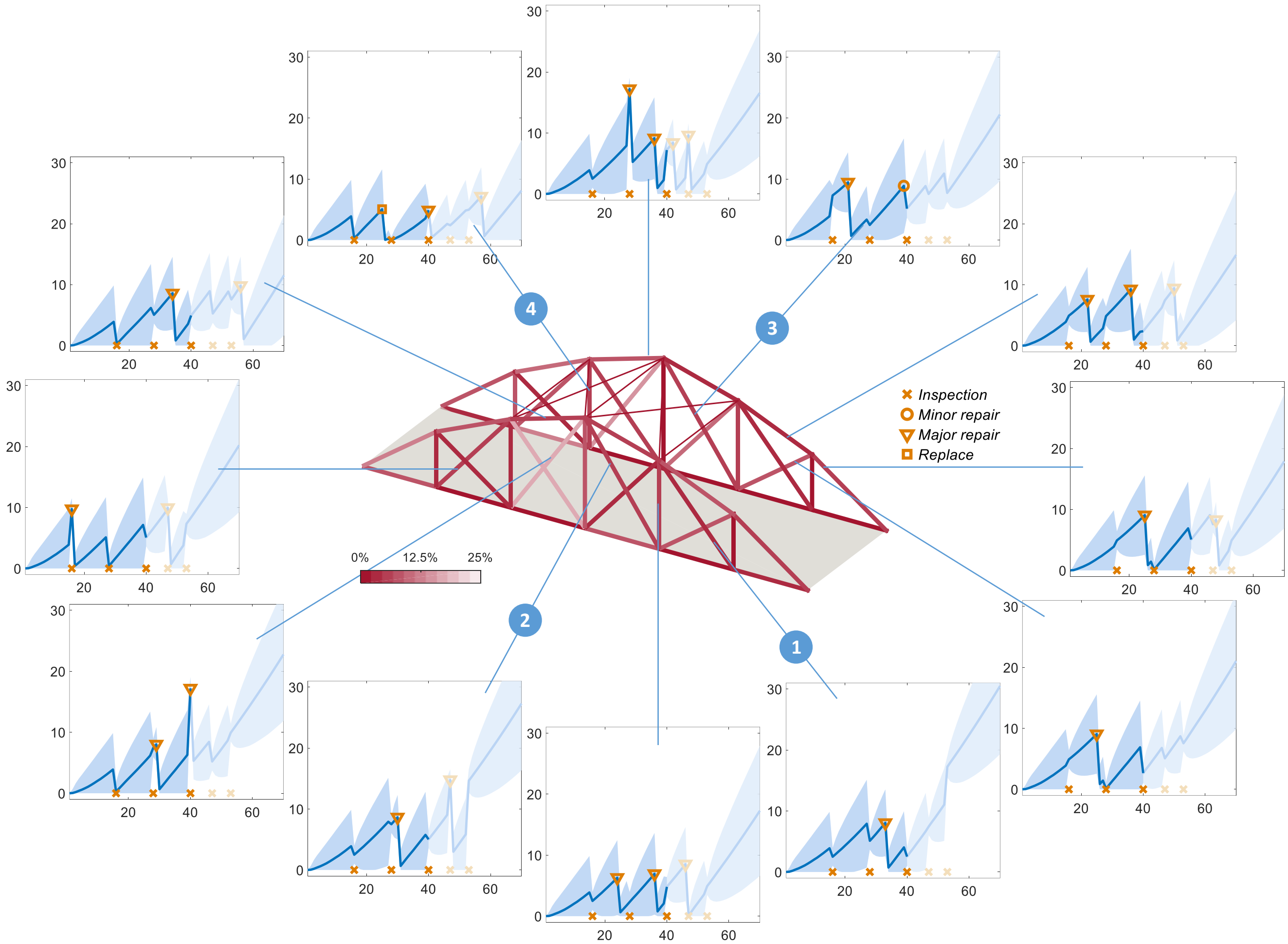

**Fig. 11.** Policy realization of System III with snapshot at time $t$ = 40 years. The structural system snapshot shows the observed section loss. Plots show the mean section loss evolution (%), ± 2 standard deviations, for different components over the planning horizon, together with the selected maintenance actions at different time steps.

Components define separate control units with respect to maintenance actions, as mentioned, whereas in this POMDP problem realistic inspection choices for the entire structure are also available at each decision step. An inspection action determines whether to inspect the bridge structure (and all its components) or not, as opposed to the previous POMDP setting in System II where inspections are permanent. This inspection choice renders the structure, as a whole, a separate control unit with respect to the observation actions. The accuracy of observations is $p = 0.90$, defined similarly as in System II, and as further explained in Appendix B. Again, there are 4 maintenance actions available per component, i.e. *do nothing* (action 1), *minor repair* (action 2), *major repair* (action 3) and *replace* (action 4). Action effects are slightly different in this setting to comply with the nature of a steel structural system. Minor repairs merely change the deterioration rate of a component by 5 years. Such actions can reflect cleaning and repainting of member surfaces, or repairs of the coating of steel structures, which do not change the damage state of the system but delay the aging process. Major repairs have the same effect as minor repairs, but in addition they also improve the state of the corresponding component by 5 states at that decision step and before the next transition. Such actions can refer to cleaning and repainting of corroded surfaces, combined with structural members strengthening. Replace indicates again at that decision step a transition to state 1 and the initial deterioration rate, as in the previous example. The total number of system states in this finite horizon example is approximately equal to $1.2 \times 10^{81}$, when all combinations of the 70 different deterioration rates and the 25 possible damage states of the 25 components are considered. Similarly, the total number of available system actions is more than $2.25 \times 10^{15}$, prohibiting again DQN approaches, or use of any other available planning algorithm, without imposing significant, and often unrealistic, simplifications, or acceding to outputs of low decision policy fidelity.

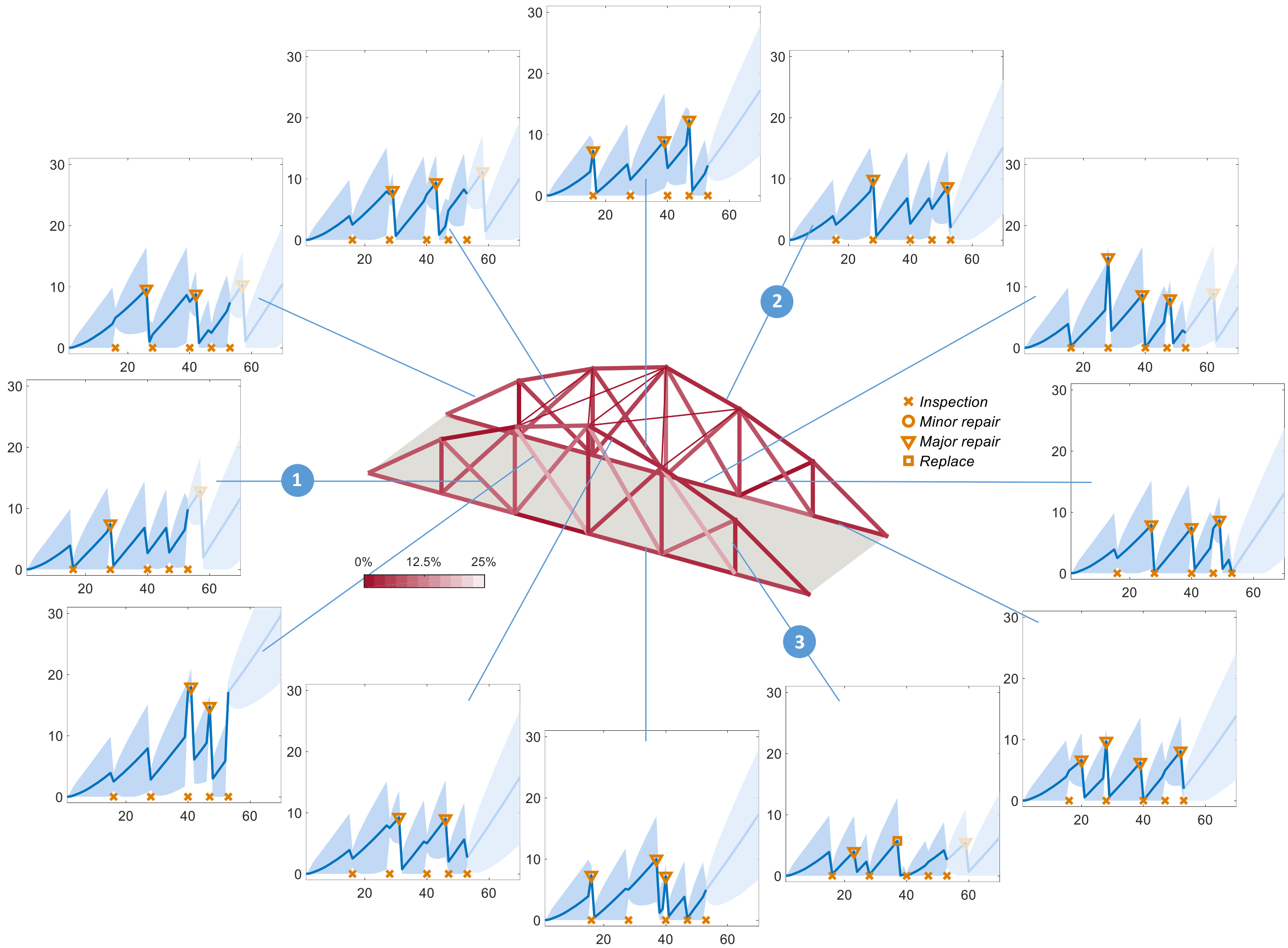

**Fig. 12.** Policy realization of System III with snapshot at time $t$ = 53 years. The structural system snapshot shows the observed section loss. Plots show the mean section loss evolution (%), ± 2 standard deviations, for different components over the planning horizon, together with the selected maintenance actions at different time steps.

All DCMAC implementation details and deep network specifications for this high-dimensional setting are discussed in Appendix C. The resulting DCMAC policies are again evaluated here against 4 baseline policies. For all 4 baselines, actions are taken based on thresholds related to the deterioration rate (for action 2) and the mean section loss (for actions 3 and 4) of each component, as this is determined based on the acquired probabilities by the Bayesian updates in Eq. (6). Each baseline has been optimized for periodic inspections every 2, 5, 10, and 15 years and the optimal state and rate thresholds for actions 2, 3 and 4 are considered uniform for all components. The resulting policies can be reviewed in Table 2, along with their respective life-cycle costs, which are determined based on $10^3$ Monte Carlo 70-year policy realizations, and are normalized with respect to the DCMAC results from the best run. Minor repair actions are not shown in the table, since their optimal deterioration rate threshold values are near the final length of the 70-year planning horizon, essentially making them redundant. As seen in Table 2, DCMAC outperforms the optimized baselines with the fixed inspection periodicity, offering life-cycle cost reductions from 5.8% to 22.3%. Training results for the best 4 DCMAC runs are shown in Fig. 10, where baseline values are also indicated on the plotted line of the best run. The worst baseline, CBM-I, is surpassed after almost 7.9 thousand training episodes, the best one, CBM-III, after 17.8 thousand episodes, whereas CBM-II and CBM-IV after 15.3 and 16.9 thousand episodes, respectively. Since Fig. 10 only displays the first 10.0 thousand episodes and the 53.0-55.0 thousand ones, for better visualization purposes, the CBM-II to CBM-IV values are shown at the breaking line of the x-axis. A well-converged near-optimal solution is practically attained after approximately 25.0 thousand episodes.

In Figs. 11 and 12, one DCMAC policy realization is visualized, emphasizing in two different time steps. In this specific realization, 5 inspection visits are chosen over the entire life-cycle of the structure at 16, 28, 40, 47 and 53 years. At the bridge state snapshots, the observed losses of all truss components are illustrated at times of 40 and 53 years, whereas the plots indicate the corresponding mean section loss evolutions (%) ± 2 standard deviations, over the planning horizon, for different components. The future mean section losses and actions are shown with lighter color. A general pattern, also present in this realization, is that inspection visits are driven primarily by increased levels of uncertainty regarding the damage state of the system. As expected, the system state uncertainty reduces when inspections are taken, even for components where no other action has been applied. However, as a result of inspections, the presumed damage state of the components can either increase or reduce, depending on the actually observed state, which can change the damage state beliefs drastically, given that the observation accuracy is high in this case ( $p = 0.90$ ). Relevant examples of damage state jumps at inspection times can be seen in Fig. 11, at $t$=28, 53, for components 1, 2 respectively, and in Fig. 12, at $t$=16, 40 for components 1, 2, respectively.

Overall, the agent favors major repair actions in this setting, since they combine features of both minor repairs and replacements. Major repairs always have a positive impact on the damage state severity, while also reducing the deterioration rate, which is translated in slope reductions of the mean section loss curves in Figs. 11-12. Consistent with the baseline policies findings, minor repairs are less frequently chosen by DCMAC, as for example in Fig. 11 at $t$=40 for component 3, as major repairs comprise their deterioration rate reduction effect. When a minor repair is performed, uncertainty and damage state remain unchanged and only the deterioration rate decreases, thus this action merely controls the non-stationarity of the system. Replacement signifies a restart for the deterioration process of a component, thus obviously reducing uncertainty, expected section loss, and deterioration rate drastically, as for example can be observed in Fig. 11 at $t$=25, for component 4, and in Fig. 12 at $t$=37, for component 3. Generally, as observed, DCMAC provides again an extremely refined and sophisticated life-cycle system policy, consisting of detailed component-level and sub-system level decisions.

## 6. Conclusions

A framework for addressing the important problem of scheduling comprehensive maintenance and inspection policies for large engineering systems is presented in this paper. Sequential decision-making over the life-cycle of a system is formulated within MDP and POMDP conceptions, which follow efficient dynamic programming principles to describe effective real-time actions under complete and partial observability. The curse of dimensionality in relation to vast state and action spaces describing large multi-component engineering systems can, however, considerably impede proper solutions in these settings. In this work, the Deep Centralized Multi-agent Actor Critic (DCMAC) approach is introduced, which constitutes the first Deep Reinforcement Learning (DRL) implementation for large multi-component engineering systems control. DCMAC has the capacity to handle immense state and action spaces and to suggest competent near-optimal solutions to otherwise intractable learning problems. DCMAC uses a centralized value function and a centralized actor network with a factorized output probability distribution representation. This architecture establishes suitable conditional independences related to component actions, making actor output dimensions to scale linearly with the number of control units, without any loss of accuracy. We present DCMAC in relation to already available DRL methods, such as DQNs and deep policy gradient approaches, which we also describe accordingly, along with their strengths, limitations and implementation details. We deploy our approach in various problem settings, featuring generic deteriorating systems and structural applications, verifying and validating that DCMAC is able to find intelligent solutions and complex decision policies in challenging high-dimensional domains, and to outperform optimized baselines that reflect standard current engineering practice. Overall, DRL under the DCMAC approach, provides unparalleled capabilities and solutions for optimal sequential decision-making in complex, non-stationary, multi-state, multi-component, partially or fully observable stochastic engineering environments, enabling several new directions for controlling large-scale engineering domains and generic multi-agent environments.

## Acknowledgements

This material is based upon work supported by the National Science Foundation under CAREER Grant No. 1751941.

## Appendix A. DQN Algorithm

DQN is implemented in its double Q-learning version (DDQN), as explained in the main text and shown in Algorithm A1. Notation here refers to MDP environments since DDQN is only utilized in this paper in the MDP setting of System I, where complete observability is considered. However, partial observability and belief MDPs can be straightforwardly applied as in Algorithm 1, presented in Section 4.

**Algorithm A1** Double Deep Q-Network (DDQN) [53].

```
Initialize replay buffer
Initialize network weights θ^Q
Initialize target network weights θ^{Q⁻}
Set target update time
for episode = 1,M do
    for t=1,T do
        Select action a_t at random according to exploration noise
        Otherwise a_t = arg max Q(s_t, a_t | θ^Q)
        Collect reward r(s_t, a_t), observe new state s_{t+1}
        Store experience tuple (s_t, a_t, r(s_t, a_t), s_{t+1}) to replay buffer
        Sample batch of tuples (s_i, a_i, r(s_i, a_i), s_{i+1}) from replay buffer
        If s_{i+1} is terminal state y_i = r(s_i, a_i)
        Otherwise y_t = r(s_i, a_i) + γQ⁻(s_{i+1}, arg max Q(s_{i+1}, a_{i+1} | θ^Q) | θ^{Q⁻})
        Update parameters according to gradient:
            g_{θ^Q} ≃ Σ_i (Q(s_t, a_t | θ^Q) − y_t) ∇_{θ^Q} Q(s_t, a_t | θ^Q)
        If target update time reached set θ^{Q⁻} = θ^Q
    end for
end for
```

## Appendix B. Environments

### B.1. Systems I and II

Environments of Systems I and II share similar specifications. The 5 components of System I and the first 5 components of System II are alike, except that the latter have non-stationary transitions. The respective damage state and maintenance costs are also identical. Transition dynamics between consecutive damage states, $x_t^{(l)}, x_{t+1}^{(l)} \in X = \{1,2,3,4\}$ have the following generic matrix form for every deterioration rate, $\tau_t^{(l)} = \tau$, and for all components $l$:

$$\mathbf{P} = \left[ p\left(x_{t+1}^{(l)} = j \mid x_t^{(l)} = i, \tau_t^{(l)} = \tau\right) \right]_{i,j\in X} = \begin{bmatrix} p_{\tau,11} & p_{\tau,12} & p_{\tau,13} & \\ & p_{\tau,22} & p_{\tau,23} & p_{\tau,24} \\ & & p_{\tau,33} & p_{\tau,34} \\ & & & p_{\tau,44} \end{bmatrix} \quad \text{(B1)}$$

Transitions are non-stationary only in System II, namely the matrix arguments of Eq. (B1), $p_{\tau,ij}$, also depend on the deterioration rate. For each deterioration rate, transitions matrices are obtained by interpolating one initial and one final transition matrix, similarly to [18] and without any loss of generality. For all components, $p_{\tau,ij}$'s are shown in Fig. B1 as functions of the deterioration rate, whereas the stationary transition probabilities for System I are defined for $\tau = 0$. For the $p_{\tau,23}$ probability entries in Fig. B1, lines of components 1 and 3 coincide. Damage state transition matrices for the different actions in System II, including both the effects of maintenance and environment deterioration for the $t$+1 state, are:

$$\mathbf{P}_1 = \left[ p\left(x_{t+1}^{(l)} = j \mid x_t^{(l)} = i, a_t^{(l)} = 1, \tau_t^{(l)} = \tau\right) \right]_{i,j\in X} = \mathbf{P},$$

$$\mathbf{P}_2 = \left[ p\left(x_{t+1}^{(l)} = j \mid x_t^{(l)} = i, a_t^{(l)} = 2, \tau_t^{(l)} = \tau\right) \right]_{i,j\in X} = \begin{bmatrix} p_{\tau,11} & p_{\tau,12} & p_{\tau,13} & \\ p_{\tau,11} & p_{\tau,12} & p_{\tau,13} & \\ & p_{\tau,22} & p_{\tau,23} & p_{\tau,24} \\ & & p_{\tau,33} & p_{\tau,34} \end{bmatrix},$$

$$\mathbf{P}_3 = \left[ p\left(x_{t+1}^{(l)} = j \mid x_t^{(l)} = i, a_t^{(l)} = 3, \tau_t^{(l)} = \tau\right) \right]_{i,j\in X} = \begin{bmatrix} p_{\tau',11} & p_{\tau',12} & p_{\tau',13} & \\ p_{\tau',11} & p_{\tau',1,2} & p_{\tau',13} & \\ & p_{\tau',22} & p_{\tau',23} & p_{\tau',24} \\ & & p_{\tau',33} & p_{\tau',34} \end{bmatrix}, \quad \text{(B2)}$$

$$\mathbf{P}_4 = \left[ p\left(x_{t+1}^{(l)} = j \mid x_t^{(l)} = i, a_t^{(l)} = 4, \tau_t^{(l)} = \tau\right) \right]_{i,j\in X} = \begin{bmatrix} p_{0,11} & p_{0,12} & p_{0,13} & \\ p_{0,11} & p_{0,12} & p_{0,13} & \\ p_{0,11} & p_{0,12} & p_{0,13} & \\ p_{0,11} & p_{0,12} & p_{0,13} & \end{bmatrix}$$

where $\tau' = \max\{0, \tau - 5\}$. Matrix entries of Eq. (B2) are given by the corresponding entries of Eq. (B1). For example, $p_{\tau,33}$ in $\mathbf{P}_2$ describes the case of a component starting at damage state 4 at $t$ and ending up at damage state 3 at $t$+1, following the relevant $p_{\tau,33}$ probabilities in Fig. B1. Actions 2 and 3 have the same effect on component damage states, as seen in Eq. (B2), while action 3 also reduces the deterioration rate by 5 steps. Both actions have a success rate of 0.95, meaning that with probability 0.05 the environment transition follows Eq. (B1). Each component damage state is associated with certain direct costs, which are essentially direct losses typically quantifying economic impact as a result of state degradation. Similarly, additional costs apply depending on the chosen maintenance actions for each component. Direct losses and maintenance costs vary between components, as shown in Fig. B2 in arbitrary units. In the direct losses plot of Fig. B2(a), lines of components 3, 6, and 9 coincide. Observation matrices for component $l$ are defined as:

$$\mathbf{O} = \left[ p\left(o_{t+1}^{(l)} = j \mid x_{t+1}^{(l)} = i, \mathbf{a}_t\right) \right]_{i,j\in X} = \begin{bmatrix} p & 1-p & & \\ (1-p)/2 & p & (1-p)/2 & \\ & (1-p)/2 & p & (1-p)/2 \\ & & 1-p & p \end{bmatrix} \quad \text{(B3)}$$

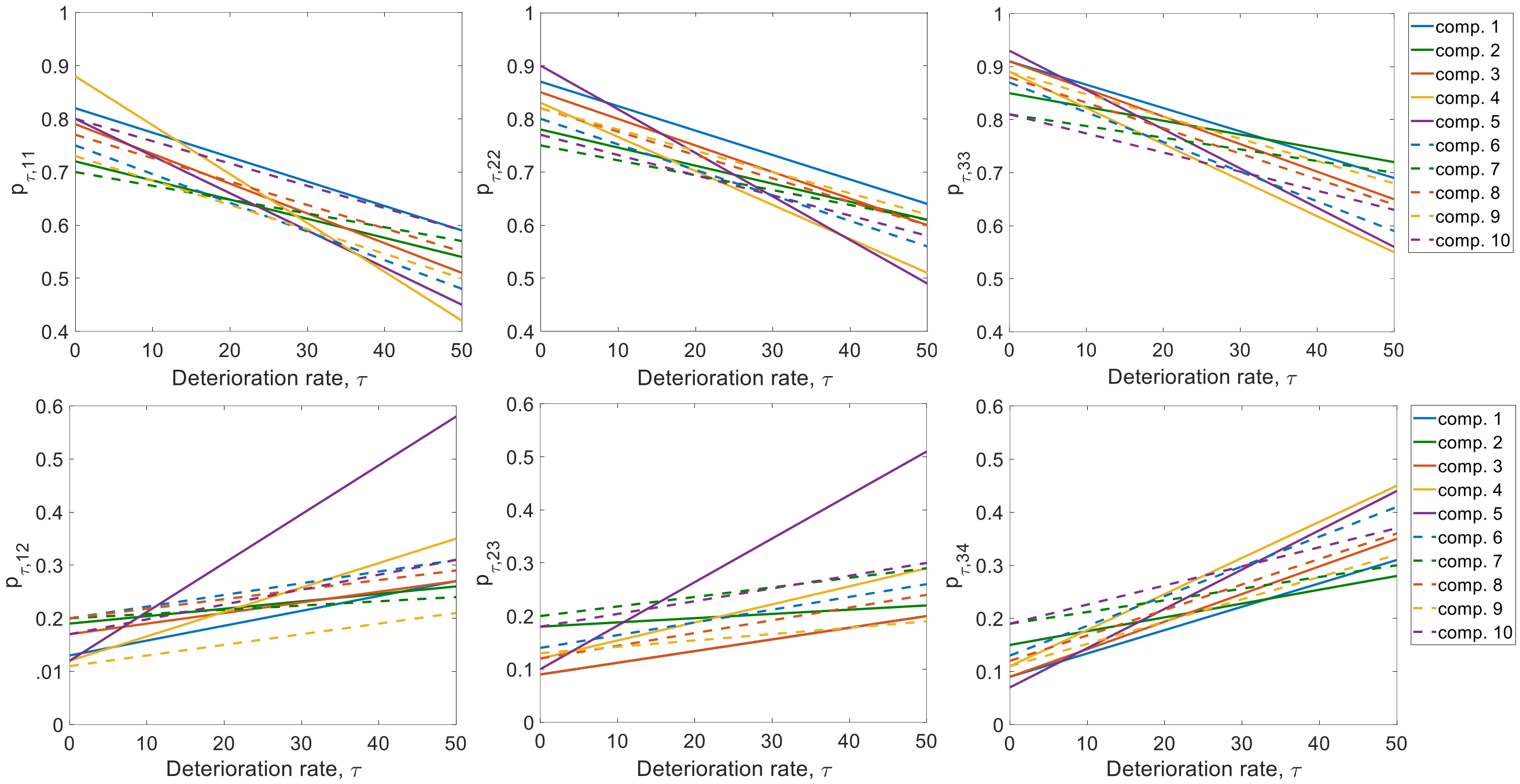

**Fig. B1.** Non-stationary transition probabilities between component damage states for System II. Plots fully define Eq. (B1) as probabilities sum to 1 for each row.

where $p$ is a given level of precision, which is constant for all available actions, $\mathbf{a}_t$. The observation matrix of Eq. (B3), suggests that the agent observes the correct component state with probability $p$, however there is always possibility of observing adjacent states. In the case of perfect observations, $\mathbf{O} = \left[ p\left(o_{t+1}^{(l)} = j \mid x_{t+1}^{(l)} = i\right)\right]$ becomes the identity matrix and Eq. (6) assigns probability 1 to the true component state.

### B.2. System III

The effect of corrosion on steel structures can be effectively described by means of cross section losses of structural members. The depth of penetration is typically considered to be proportional to a power law of exposure time [1, 72]. Assuming hollow cross sections of same thickness for all members, the mean uniform section loss percentage, $d_m$, can be similarly approximated as:

$$d_m = \frac{A_\tau^{cs}}{A_0^{cs}} \propto \tau^\beta \tag{B4}$$

where $A_\tau^{cs}$ is the member cross section area at exposure time $\tau$, and $\beta$ is a constant with typical values between $0 < \beta \le 2.0$, depending on environmental conditions, corrosion causes, material properties, etc. To model the stochastic nature of the phenomenon, we consider the section loss percentage, $d$, as a gamma process with mean value $d_m$. Exposure time is assumed to define here the value of deterioration rate, e.g. a member with 3 years of exposure (without any maintenance) is at deterioration rate 3. The marginal probability of $d$ at every step $\tau$ follows a gamma distribution with probability density function:

$$Ga\left(d \mid f(\tau), \lambda\right) = \frac{\lambda^{f(\tau)}}{\Gamma\left(f(\tau)\right)} d^{f(\tau)-1} e^{-\lambda d} I\left\{d \ge 0\right\} \tag{B5}$$

where $\lambda$ is a scale parameter, $f$ is a non-negative shape function of exposure time, $\Gamma(u) = \int_0^\infty v^{u-1} e^{-v} dv$, and $I\left\{d \ge 0\right\}$ is an indicator function assuring positivity of $d$. The mean value and standard deviation of this gamma distribution in time are:

$$d_m = \frac{f(\tau)}{\lambda}, \ \sigma = \frac{\sqrt{f(\tau)}}{\lambda} \tag{B6}$$

Gamma processes have been shown to be a suitable modeling choice in numerous stochastic deterioration engineering applications [73], and can readily describe continuous Markovian transitions in discrete time steps. That is, for two time steps, $\tau_1 < \tau_2$, the increment of $d$ also follows a gamma distribution:

$$d(\tau_2) - d(\tau_1) \sim Ga\left(\cdot \mid f(\tau_2) - f(\tau_1), \lambda\right) \tag{B7}$$

It can be seen through Eqs. (B4) and (B7) that for $\beta = 1.0$ the gamma process is stationary, meaning that it only depends on the time distance between $\tau_1, \tau_2$. In this example, $\beta = 1.5$ is used, which is indicative of non-stationary damage evolution that is anticipated in highly corrosive environments. To determine the scale and shape parameters of the gamma process, we assume a mean section loss of 40% with a standard deviation of 7.5% in 70 years, which approximately corresponds to a 4mm corrosion penetration, on average, at the end of the life-cycle horizon. The resulting deterioration process is illustrated in Fig. B3. The underlying conti-nuous space is properly discretized here with a step of 2.5%, to support standard discrete state Markovian transitions. We also assume that if the section loss of a member exceeds 60%, that member is considered failed, hence the total number of damage states per component is 25. This results in 25x25 transition matrices for each deterioration rate, which have been estimated by $10^6$ Monte Carlo simulations of the gamma process.

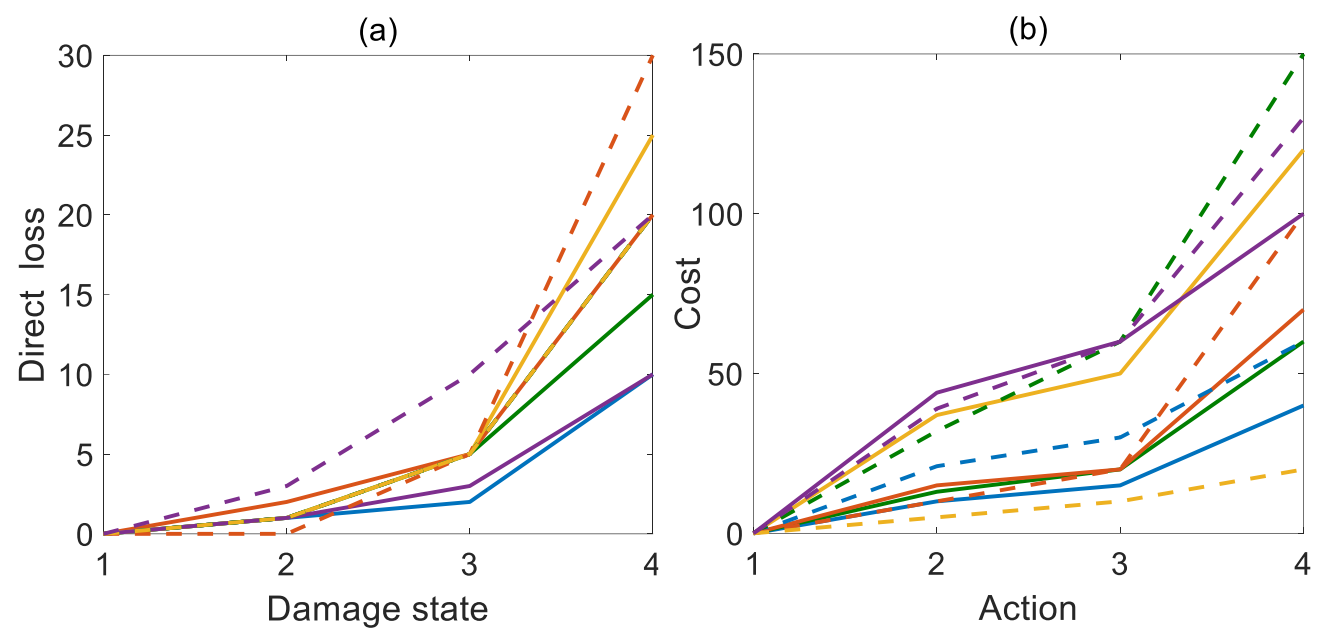

**Fig. B2.** (a) Direct loss related to damage states and (b) cost of maintenance actions, for all components (Systems I and II). Colors indicate different components according to Fig. B1.

The corresponding component transition probabilities given an action have been calculated as previously, with proper adjustments concerning the particular action effects in this problem, as explained in the main text. Direct losses as a result of component deterioration are considered to be proportional to steel volume losses, with a maximum value of 30, when a component fails. Maintenance costs are likewise proportional to the initial volume of the member, with maximum values of 0, 25, 60 and 130 for actions 1 to 4, respectively. Observation matrices follow the same logic as in Eq. B3 for a given level of precision *p*. Observations are not available in this example, unless an inspection action that incurs a cost of 50 is selected, covering inspections for all components.

## Appendix C. Deep networks

Various deep network architectures have been examined, regarding the depth, the dimensions and activations of the hidden units. The networks implementations that have been finally employed in this work are fully described in this appendix section. For all examples, the deep learning Python libraries of Keras with Tensorflow backend have been utilized [74, 75].

For all examples, input has the same format and consists of the global time index of the finite horizon, *t*, and the states of all components *l*, defined as $s_t^{(l)} = \left(x_t^{(l)}, \tau_t^{(l)}\right)$, which include their damage states and deterioration rates (for non-stationary problems). Component damage states are transformed to a one-zero vectorized representation. This means that a component damage state is described by a zero vector, having only one at the entry associated with its state. In the case of partial observability, this vector does not only consist of one-zero arguments, but it is rather a vector of real numbers describing a probability distribution over all component damage states. Deterioration rates and time input are normalized with respect to the length of the planning horizon, so that their relevant values lie in $[0,1]$. Input scaling is critical in deep neural networks training, and normalization helps in an overall more stable algorithmic performance.

In System I, DQN is trained with two fully connected layers of

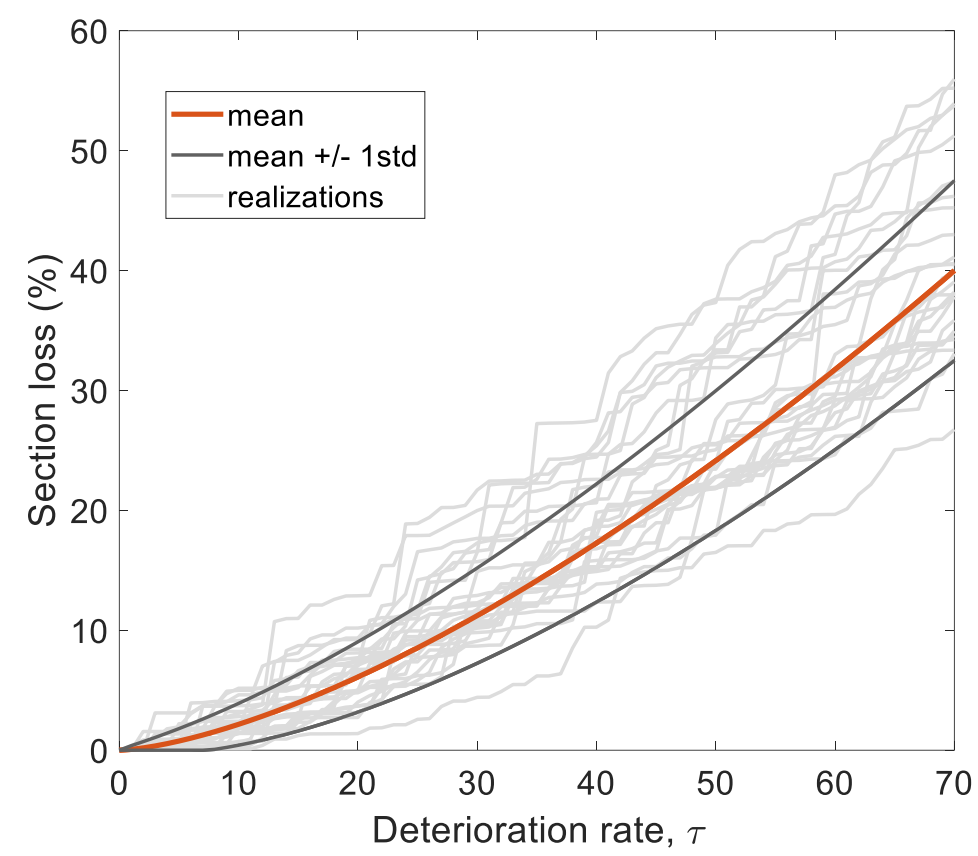

**Fig. B3.** Non-stationary gamma process describing steel cross section loss due to corrosion, over the planning horizon of 70 years, without any maintenance action effects (System III).

40 dimensions each (40x40), whereas the same hidden layer architecture is implemented for both the actor and critic networks of DCMAC. Hidden layers have been modeled in all networks using Rectified Linear Units (ReLU), which are popular in deep learning applications due to their robust training behavior and sparse activations [76]. The output layer of DQN has linear activations that approximate the 32 Q-functions related to all available actions in that problem. In the implementation of DCMAC, the actor network output consists of 5 softmax functions, as many as the number of control units in this case, providing binary values, related to the 2 available actions per component, whereas the critic network has a single linear output approximating the value function. For Systems II and III, actors and critics also have two fully connected ReLU layers, 100x100 and 350x350, respectively, whereas critic outputs are again single linear units that provide an approximate representation of the value function. The actor output for System II consists of 10 softmax functions with 4 dimensions each, which represent the available actions for each control unit. For the actor network of System III, except for the accordingly defined 25 softmax outputs of length 4, there is also a binary softmax output indicating whether the inspection action is taken or not.

Parametric updates are executed through batch training, thus using sample-based estimates of the actual gradients. This concept outlines the notion of stochastic gradient descent methods and is central in deep learning applications, as it facilitates faster training while maintaining nice convergence properties. A number of available first order optimization algorithms exists that rely on stochastic gradients to approach local minima, incorporating various enhancing concepts and heuristics, like momentum, adapting learning rate rules, subgradient principles and Nesterov steps, among others [77, 78, 79]. The Adam optimizer is used herein, which is an advanced stochastic gradient descent method, shown to have robust performance in several applications [78].

Learning rates ranging from $10^{-4}$ to $10^{-5}$ and $10^{-3}$ to $10^{-4}$ have been used for the actor and critic networks, respectively, in all examples. Starting from higher values, learning rates were being adjusted during training, up to their lowest values, to ensure smoother training. A batch size of 32 has been used, being sampled from a buffer memory size of $2\text{-}3\cdot10^5$. In DCMAC, importance sampling weights have been truncated at $c=2$, to reduce variance of the gradient estimator, as explained in the main text. Target networks, which have only been used in the DQN implementation, were updated every 13 steps, whereas an exploration noise of at most 1% was maintained throughout training of all networks, typically starting from values near 100%. Especially for System III, exploration at the early training stages with a prioritized *do nothing* action has been noticed to improve convergence and stability of the overall learning progress, whereas this was also the case for a small penalization of certain a priori known sub-optimal decisions, e.g. action 2 at the initial deterioration rate. Overall, different hyperparameters are expected to moderately affect the training time until convergence and, in some cases, the quality of the converged solution. However, suggestion of highly "optimized" architectures and hyperparameters is beyond the scope and generality of this work. Although the final specifications of our examples come as a result of thorough experimentation, comparable solutions can be potentially attained with finer tuning of network specifications